\documentclass[pra,twocolumn,a4paper,aps,10pt]{revtex4-2}

\usepackage{graphicx}
\usepackage{amssymb}
\usepackage{bm}
\usepackage[amssymb,thickqspace]{SIunits}
\usepackage{physics}
\DeclareMathOperator{\ttr}{tr}
\DeclareMathOperator{\sgn}{sgn}
\DeclareMathOperator{\texp}{\exponential}
\newcommand{\cc}{\text{c.c.}}
\newcommand{\Hc}{\text{H.c.}}
\newcommand{\up}{{\uparrow}}
\newcommand{\dn}{{\downarrow}}
\renewcommand{\Re}{\operatorname{Re}}
\renewcommand{\Im}{\operatorname{Im}}
\newcommand{\lb}{\linebreak[1]}
\begin{document}

\title{Optical Schrödinger cat states generated using cavity Rydberg quantum photonics}
\author{Hendrik Hegels}
\author{Mae Eichenberger}
\author{Gerhard Rempe}
\author{Stephan D\"urr}
\affiliation{Max-Planck-Institut f\"{u}r Quantenoptik, Hans-Kopfermann-Stra{\ss}e 1, 85748 Garching, Germany and \linebreak[3] Munich Center for Quantum Science and Technology, Schellingstra{\ss}e 4, 80799 M\"{u}nchen, Germany}

\begin{abstract}
We introduce cavity Rydberg quantum photonics as a novel platform for generating hybrid entangled states of optical photons. The system achieves world-record sizes of these states and has the potential to produce yet larger states in the future. In the hybrid entangled state produced here, a superposition of two coherent states is entangled with a polarization qubit carried by a photon in another beam path. We perform quantum state tomography of this hybrid entangled state and evaluate various key properties. By optimizing the decoherence we increase the size compared to previous hybrid entangled states, reaching a size of $\alpha^2= 2.4$. Besides the fundamental interest in Schrödinger cat states, this work is interesting because of potential applications in the field of quantum information processing.
\end{abstract}

\maketitle

\textit{Introduction} --
If a quantum superposition of macroscopically different states is additionally entangled with a microscopic degree of freedom, the composite state is referred to as a Schrödinger cat state \cite{Schroedinger:35:cat}. The experimental generation of Schrödinger cat states is a topic of intense contemporary research \cite{Monroe:96, Auffeves:03, Lo:15, Wang:22}. In some parts of the literature, the name Schrödinger cat state is also used to refer to superpositions of macroscopically different states in the absence of entanglement \cite{Friedman:00, Kovachy:15, Fein:19, Milul:23, Bild:23} or to Greenberger-Horne-Zeilinger states \cite{Omran:19, Song:19, Pogorelov:21, Thomas:22}. Exploring all of the above states in different physical systems and pushing towards more macroscopic superpositions is interesting because of fundamental questions, e.g., whether there is a spontaneous collapse of macroscopic superpositions \cite{Bassi:13}. In addition, this research is motivated by a broad spectrum of potential applications in quantum sensing, quantum metrology, and quantum information processing \cite{Ralph:03, Gilchrist:04, Giovannetti:11, Degen:17}.

For optical photons, superpositions of macroscopically different states in the absence of entanglement are typically implemented as coherent state superpositions (CSSs), i.e., superpositions of coherent states $\ket\alpha$ and $\ket{-\alpha}$ \cite{Takahashi:08, Gerrits:10, Lewenstein:21} or squeezed versions thereof \cite{Ourjoumtsev:07, Huang:15, Etesse:15, Ulanov:16, Chen:24, Simon:24, Caron:26}. Entangling such a CSS of optical photons with a discrete variable (DV) generates a class of Schrödinger cat states, which is called hybrid entangled states (HESs). The latter are particularly interesting because they are a crucial resource for a hybrid approach to optical quantum computing and quantum communication \cite{vanLoock:11, Lee:13}. While the two traditional approaches rely on either DVs \cite{Knill:01} or continuous variables \cite{Jeong:02, Ralph:03}, which both have their advantages and disadvantages, the hybrid approach combines these two encodings and can profit from the strengths of both implementations. This stimulated a considerable amount of work on the theory of this hybrid approach \cite{vanLoock:11, Lee:13}. Such HESs have been generated using traditional nonlinear media combined with postselection \cite{Morin:14, Jeong:14, Takeda:15, Xu:25} and a few simple applications have already been demonstrated \cite{Ulanov:17, Sychev:18, LeJeannic:18, Darras:23}. While CSSs with $|\alpha|^2= 3.1$ and squeezed CSSs with $|\alpha|^2= 6$ have been demonstrated, the generation of HESs using traditional nonlinear media only reached $|\alpha|^2=0.8$ \cite{Gerrits:10, Caron:26, Morin:14}. For some applications, however, it is necessary to achieve HESs with larger $\alpha$ \cite{vanLoock:11, Lee:13}.

This raises the question whether quantum systems that exhibit a large nonlinearity down to the level of single photons might offer a path to HESs with larger $\alpha$. Indeed, there is a single experiment that generated HESs with a quantum nonlinear system, namely using a single atom in a high-finesse cavity \cite{Hacker:19}. This experiment outperformed the traditional approach and reached $|\alpha|^2=1.4$. The dominant limitation of that experiment was decoherence caused by the which-way information carried away by photons lost from the cavity during state generation. Recently, we proposed \cite{Hegels:26:theo} that this problem can be mitigated considerably when using cavity Rydberg electromagnetically induced transparency (EIT) because this makes it possible to evade the leakage of which-way information into the environment even if the photon loss is not small \cite{Hegels:26:theo}. Rydberg EIT relies on Rydberg interactions, which can mediate nonlinear optics at the single-photon level \cite{Firstenberg:16}. This has led to studies of a large variety of phenomena \cite{Pritchard:10, Peyronel:12, Dudin:12, Firstenberg:13, Gorniaczyk:14, Tiarks:14, Thompson:17, Tiarks:19, Stiesdal:21, Lee:23, Sumarac:2601.06345}. Combining Rydberg EIT with an optical cavity enlarged the capabilities in the field even further \cite{Schine:16, Clark:20, Stolz:22, Magro:23, Aggarwal:2602.18363}.

Here, we introduce cavity Rydberg EIT as a novel platform for generating HESs. We achieve a world-record size of $|\alpha|^2 = 2.4$ of optical HESs. Performing an appropriate measurement on the DV and postselecting upon the measurement outcome, we also obtain CSSs, which are no longer entangled with the DV. We perform quantum state tomography of the generated HESs and CSSs and evaluate various key properties. As expected \cite{Hegels:26:theo}, the decoherence during state generation is reduced compared to Ref.\ \cite{Hacker:19}, leading to improved performance in these key properties. An analysis of the present limitation suggests that the system has the potential to produce yet larger states in the future. Theses results show that cavity Rydberg EIT is an attractive path for taking HESs to larger $\alpha$, a key ingredient needed for putting HESs to use in many quantum information applications.

\begin{figure}[tb!]
\centering
\includegraphics[scale=1]{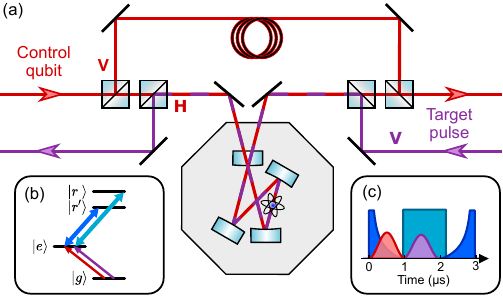}
\caption{(a) Scheme of the cat-state generation. A polarization qubit carried by the control photon (red) is converted into a dual-rail qubit by a polarizing beam splitter (PBS, blue square). In one rail the photon is stored as a Rydberg excitation inside the cavity, in the other rail the photon bypasses the cavity. Next, the target light pulse (purple) enters the setup and experiences a conditional $\pi$ phase shift upon reflection off the cavity. Subsequently, the control photon is retrieved and overlapped with the bypass on another PBS. An optical fiber in the bypass matches the delay resulting from storage. The vacuum chamber is represented by the octagon. The scheme is simplified for clarity. In particular, waveplates are not shown. (b) Atomic level scheme. Coupling light (blue, cyan) creates Rydberg EIT for the signal light (red, purple). (c) Timing sequence of the light power of the incoming pulses. First, the control pulse is stored, second, the target pulse is reflected off the cavity, and, third, the control photon is retrieved.}
\label{fig-scheme}
\end{figure}

\textit{Cat-state generation} --
Following the proposals in Refs.\ \cite{Wang:05:cat, Hegels:26:theo}, we generate HESs in the setup described in detail in Ref.\ \cite{Stolz:22}, see Fig.\ \ref{fig-scheme}. In contrast to Ref.\ \cite{Stolz:22}, the incoming and outgoing target light pulse is always vertically polarized and it is in a coherent state with variable amplitude $\alpha_\text{in}$.

If the incoming control photon has horizontal (vertical) polarization, $\ket H$ ($\ket V$), then this photon will be stored in the atomic ensemble (propagate through the fiber). However, if the incoming control photon is in the state $\frac1{\sqrt2}(\ket H + \ket V)$, then the final state should ideally be the pure HES \cite{Hegels:26:theo}
\begin{align}
\label{psi-HES}
\ket{\psi_\text{HES}} =
\frac{\ket{H,\alpha_H} + \ket{V,\alpha_V}}{\sqrt2}
,\end{align}
where $\alpha_{H/V}$ are complex amplitudes of coherent states. Note that states $\ket*{\alpha_{H/V}}$ have the same target polarization $\ket V$. Their subscript H/V refers to the control polarization they correlate with. As an aside, we can also prepare the initial state $\frac1{\sqrt2}(\ket H + e^{-i\theta_\text{ent}} \ket V)$ with an arbitrary phase $\theta_\text{ent}\in \null ]{-}\pi,\pi]$ and obtain a corresponding phase in Eq.\ \eqref{psi-HES}. The effective size of the cat state is defined as \cite{Hacker:19, Hegels:26:theo}
\begin{align}
\label{alpha-out-def}
\alpha_\text{eff}
= \frac{\abs{\alpha_H - \alpha_V}}{2}
.\end{align}
In contrast, the value of $\alpha_H + \alpha_V$ is of little interest because it could be modified easily with a displacement \cite{Leonhardt:97}.

Reference \cite{Hegels:26:theo} suggests that loss of target photons will be the dominant experimental imperfection. After tracing out the lost photons, one obtains the mixed HES \cite{Hegels:26:theo}
\begin{multline}
\label{rho-HES}
\rho_\text{mix}
= \tfrac12 \dyad{H,\alpha_H} +\tfrac12 \dyad{V,\alpha_V}
\\
+ \tfrac12 V (\dyad{H,\alpha_H}{V,\alpha_V} + \Hc)
,\end{multline}
where $V\in[0,1]$ is the visibility. $V$ is expected to depend on $\alpha_\text{eff}$ as \cite{Hegels:26:theo}
\begin{align}
\label{V-V0-LEff-alphaEff}
V
= V_0 \exp ( -2 \frac{L_\text{eff}}{1-L_\text{eff}} \alpha_\text{eff}^2 )
\end{align}
with real parameters $V_0$ and $L_\text{eff}$, which are independent of $\alpha_\text{eff}$. $L_\text{eff}$ is an effective loss coefficient. We distinguish between loss \emph{during} and \emph{after} cat-state generation. They are characterized by the effective loss coefficients $L_g$ and $L_d$ resulting from generation and detection, respectively. Combining them, we obtain the total effective loss coefficient
\begin{align}
\label{L-t}
L_t
= 1-(1-L_d)(1-L_g) 
.\end{align}
We denote the value of $\alpha_\text{eff}$ ($V$) just after the cavity as $\alpha_g$ ($V_g$) and its value recorded by the detector as $\alpha_d$ ($V_d$). Hence, Eq.\ \eqref{V-V0-LEff-alphaEff} implies $V_g= V_0 \exponential (-2 \frac{L_g}{1-L_g} \alpha_g^2 )$ and $V_d= V_0 \exponential (-2 \frac{L_t}{1-L_t} \alpha_d^2 )$ with the same $V_0$. In our setup, we measure a detection loss of $L_d= 21$\% \cite{supplemental:model:experiment}.

\begin{figure*}[t!]
\centering
\includegraphics[width=\textwidth]{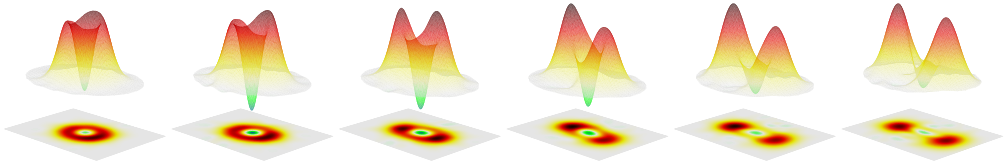}
\caption{Wigner functions of odd cat states of six different sizes. The data are not corrected for detection loss. All Wigner functions have negative regions (green), thus featuring a nonclassical property. The loss-corrected cat size increases from left ($\alpha_g^2=0.18$) to right ($\alpha_g^2=2.4$).}
\label{fig-Wigner}
\end{figure*}

\textit{Tomography of the HES} --
As in Ref.\ \cite{Stolz:22}, we use polarization optics and single-photon counters to perform a measurement on the polarization qubit of the outgoing control photon with a measurement basis, which can be chosen arbitrarily. As the measurement observables, we typically choose one of the three Pauli matrices with respect to the $(\ket H,\ket V)$ basis. Hence, the measurements have eigenstates $(\ket D,\ket A)$, $(\ket L,\ket R)$, or $(\ket H,\ket V)$. Here, $\ket D= \ket{\psi_0}$, $\ket A= \ket{\psi_\pi}$, $\ket L= \ket*{\psi_{\pi/2}}$, and $\ket R= \ket*{\psi_{3\pi/2}}$ denote diagonal ($45^\circ$), antidiagonal ($-45^\circ$), lefthand circular, and righthand circular polarization \cite{Stolz:22}, respectively, where
\begin{align}
\label{psi-theta}
\ket{\psi_\theta}
= \frac{\ket H + e^{i\theta} \ket V}{\sqrt2} 
.\end{align}
Postselecting the data upon the measurement outcome corresponding to $\ket{\psi_\theta}$, we expect to obtain the mixed CSS
\begin{align}
\label{rho-CSS}
\rho_\theta
= \frac{\dyad{\alpha_H} + \dyad{\alpha_V} + V (e^{i\theta} \dyad{\alpha_H}{\alpha_V}+\Hc)}{4\tau_\theta} 
,\end{align}
where $\tau_\theta$ is the probability of obtaining the measurement outcome $\ket{\psi_\theta}$ \cite{supplemental:model:experiment}. We use the name \emph{cat state} to refer to a HES or a CSS. For $\alpha_V= -\alpha_H$, the CSS with $\theta=0$ ($\theta=\pi$) is called an even (odd) cat state because for $V=1$ it contains only even (odd) Fock states.

We expand the experimental setup of Ref.\ \cite{Stolz:22} by a balanced homodyne detector for the outgoing target light. This makes it possible to perform homodyne tomography of the target light \cite{supplemental:model:experiment}. For reconstructing the target-light density matrix, we use the $R\rho R$ algorithm \cite{Lvovsky:04}, in which one can correct for the detection loss $L_d$, if desired \cite{Banaszek:99}. For the HES, quantum state tomography is obtained by combining the six density matrices reconstructed when postselecting upon $\ket H$, $\ket V$, $\ket D$, $\ket A$, $\ket L$, and $\ket R$ \cite{Morin:14, Hacker:19, supplemental:model:experiment}.

\textit{Wigner functions} --
Throughout the remainder of this letter, we discuss experimental results obtained for six different values of $\alpha_\text{in}$. For each of them, we perform quantum state tomography to reconstruct the HES. To illustrate at least a fraction of the results, the Wigner functions $W(x,p)$ \cite{Leonhardt:97} of the odd CSSs obtained by postselection upon $\ket A$ are show in Fig.\ \ref{fig-Wigner}.

\textit{Comparison with the model} --
It is interesting to study how accurately the measured HES is modeled by the above theory. To this end, we fit the mixed HES $\rho_\text{mix}$ of Eq.\ \eqref{rho-HES} to the measured HES $\rho_\text{meas}$ and obtain best-fit values for $V$, $\alpha_H$, and $\alpha_V$. From the latter two, we calculate $\alpha_\text{eff}$. We perform this fit with and without previously correcting for the detection loss in the homodyne tomography. Among other things, the loss correction affects the resulting value of $\alpha_\text{eff}$.

\begin{figure*}[!t]
\centering
\includegraphics[width=\hsize]{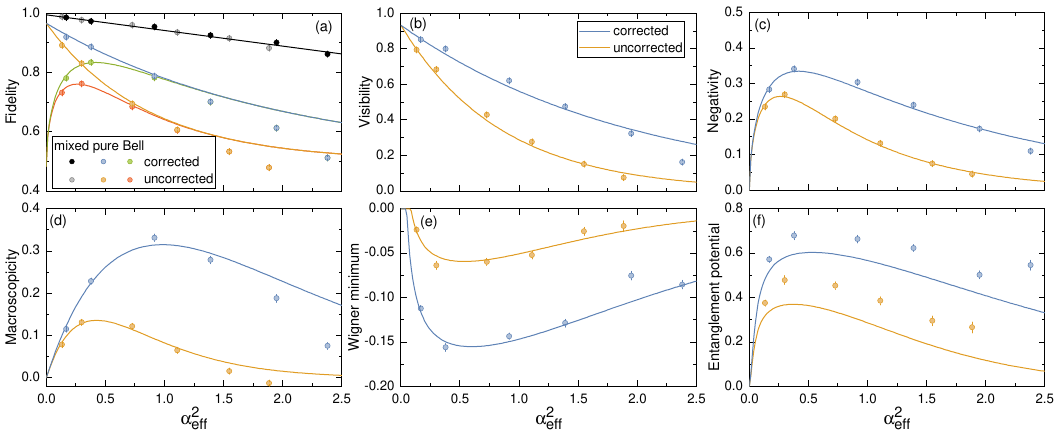}
\caption{Cat-state properties. (a) The fidelities $F_\text{pure}$ ($F_\text{mix}$) between the measured HES and the pure (mixed) best-fit HES show that the dominant limitation in the cat-state generation is a reduced visibility $V$. (b) The dependence of $V$ on $\alpha_\text{eff}^2$ agrees well with a model that assumes that loss of target photons is the dominant experimental imperfection. The Bell-state fidelity (a) and the negativity (c) show that the HES is entangled. (d) The macroscopicity measures how macroscopic the superposition in the HES is. If the minimum of the Wigner function (e) is negative or if the entanglement potential (f) is nonzero, then the odd CSS is nonclassical.}
\label{fig-entangled}
\end{figure*}

To quantify the goodness of fit, we calculate the Uhlmann fidelity \cite{Baldwin:23} $F_\text{mix} = [\tr( \sqrt{\rho_\text{meas}\rho_\text{mix}} )]^2$ between the measured HES $\rho_\text{meas}$ and the best-fit HES $\rho_\text{mix}$ of Eq.\ \eqref{rho-HES}. All results exceed 86\%, see the black and gray points in Fig.\ \ref{fig-entangled}(a).

The more interesting question is, how closely the experimental data match the pure HES \eqref{psi-HES}. The corresponding fidelity $F_\text{pure}$ is shown in Fig.\ \ref{fig-entangled}(a) as blue and orange points. As the black (gray) points lie much closer to unity than to the blue (orange) points, the discrepancy between the data and the pure state are dominantly caused by the deviation between HESs \eqref{psi-HES} and \eqref{rho-HES}, i.e., by the fact that $V<1$. Analyzing the physical origin of the much smaller deviation of the black and gray data from unity is beyond the present scope. Hence, the black straight line only serves to guide the eye.

To understand how well the dependence of the visibility $V$ on $\alpha_\text{eff}^2$ is captured by the model, we plot the best-fit values for $V$ versus $\alpha_\text{eff}^2$ in part (b) of the figure. We fit the model of Eq.\ \eqref{V-V0-LEff-alphaEff} simultaneously to \emph{all} data in Fig.\ \ref{fig-entangled}(b). In doing so, we rigidly connect $L_t$ and $L_g$ by Eq.\ \eqref{L-t} with fixed $L_d=21\%$. Hence, there are only two free fit parameters. We obtain the best-fit values $V_0=93.3(11)\%$ and $L_g=20.2(11)\%$. The fit agrees well with the data, which means that Eqs.\ \eqref{rho-HES} and \eqref{V-V0-LEff-alphaEff} capture the dominant experimental imperfections. This is consistent with loss of photons being the dominant imperfection, as expected.

We use these best-fit values to obtain a theoretical expectation for the pure-state fidelity $F_\text{pure}$, i.e, for the blue and orange points in part (a). To this end, we calculate the fidelity between HESs \eqref{psi-HES} and \eqref{rho-HES}. This gives $F_\text{pure}= \frac12(1+V)$. We insert $V$ of Eq.\ \eqref{V-V0-LEff-alphaEff} with the best-fit values for $V_0$ and $L_g$ and obtain the blue and orange lines in part (a). Below, we use the same strategy to plot expectations for other quantities. The expectation here agrees fairly well with the data. The deviation is largely caused by the fact that the measured HES deviates from the best-fit HES \eqref{rho-HES}, namely by an amount characterized by the black and gray points.

\textit{Properties of the HESs} --
We turn to evaluating three interesting properties of the HESs. First, we calculate the fidelity $F_\text{Bell}$ between the measured HES and the Bell state $\frac1{\sqrt2}(\ket{H,\up}+\ket{V,\dn})$, where $(\ket\up,\ket\dn)$ is an optimized orthonormal sequence in Fock space \cite{supplemental:model:experiment}. $F_\text{Bell}$ is interesting because $F_\text{Bell}>50\%$ is an entanglement witness, known as the Peres-Horodecki criterion \cite{Peres:96, Horodecki:96}. Results for $F_\text{Bell}$ are shown as green and red points in Fig.\ \ref{fig-entangled}(a). For large $\alpha_\text{eff}^2$, they are essentially invisible because they almost coincide with the blue and orange points. Unlike the blue and orange lines in part (a), which are decreasing functions of $\alpha_\text{eff}^2$, $F_\text{Bell}$ is non-monotonic. The measured loss-corrected data for $F_\text{Bell}$ reach a maximum of 83.4(3)\%. At $\alpha_g^2=2.4$, we measure the loss-corrected values $F_\text{Bell}=51.2(4)\%$ and $V=16.2(8)\%$. Green and red lines show the expectation \cite{supplemental:model:experiment} $F_\text{Bell}= \frac14 (1+V) \linebreak[1] [1+(1-e^{-4\alpha_\text{eff}^2})^{1/2}]$. The lack of orthogonality of states $\ket{\alpha_H}$ and $\ket{\alpha_V}$ is the reason, why $F_\text{Bell}$ deviates from $F_\text{pure}$, in particular for small $\alpha_\text{eff}^2$.

Next, we calculate the negativity $\mathcal N$ \cite{Vidal:02} of the measured HES. It is given by $\mathcal N= -\sum_i \mu_i$, where the $\mu_i$ are the negative eigenvalues of the partial transpose of the density matrix. $\mathcal N$ is interesting because it is an entanglement measure. Its range is $0\leq \mathcal N\leq \frac12(\min\{d_1,d_2\}-1)$ for a tensor product of spaces of dimensions $d_1$ and $d_2$ \cite{supplemental:model:experiment}. Here, $d_1=2$ and $0\leq\mathcal N\leq \frac12$. In Fig.\ \ref{fig-entangled}(c), the maximum loss-corrected negativity of $\mathcal N= 0.341(3)$ is not very far from the upper bound $\frac12$. At $\alpha_g^2=2.4$, the negativity is $\mathcal N= 0.110(4)$ with and 0.045(4) without loss correction. Lines show the expectation \cite{supplemental:model:experiment} $\mathcal N= \frac14\{-1+V+[(1+V)^2-4V e^{-4\alpha_\text{eff}^2}]^{1/2}\}$.

As a third property of the HES, we calculate the Lee-Jeong macroscopicity \cite{Lee:11}. It can be defined as $\mathcal I(\rho)= -\tr[\rho\mathcal L(\rho)]$ with $\mathcal L(\rho) = \hat a\rho \hat a^\dag - \frac12 \rho \hat a^\dag \hat a - \frac12 \hat a^\dag \hat a \rho$, where $\hat a$ is the photon annihilation operator. $\mathcal I$ quantifies to which degree superposed states are macroscopically different and it takes into account how coherent the superposition is. $\mathcal I$ is real and has no upper bound. $\mathcal I$ was originally introduced for quantum states of a single-mode light field, like CSSs, but generalizing it to HESs is straightforward. In Fig.\ \ref{fig-entangled}(d), the maximum loss-corrected $\mathcal I$ is 0.332(8). Lines show the expectation \cite{supplemental:model:experiment} $\mathcal I= V^2 \alpha_\text{eff}^2$.

\textit{Properties of the odd CSSs} --
We return to the odd cat states of Fig.\ \ref{fig-Wigner} and evaluate two interesting properties thereof. One property, as already mentioned, is the minimum $W_\text{min}$ of the Winger function $W$. Its range is $-\frac1\pi \leq W\leq \frac1\pi$ \cite{Leonhardt:97}. If there is a point in phase space, at which the Wigner function is negative, then this shows that the state is nonclassical \cite{Leonhardt:97}. The minima of the Wigner functions of the odd cat states of Fig.\ \ref{fig-Wigner} are shown as points in Fig.\ \ref{fig-entangled}(e). The smallest loss-corrected $W_\text{min}$ is $-0.156(4)$. At $\alpha_g^2= 2.4$, we find a loss-corrected $W_\text{min}=-0.085(5)$. The lines show the expectation \cite{supplemental:model:experiment} $W_\text{min} = \frac1\pi (e^{-2\alpha_\text{eff}^2}-V) / (1-Ve^{-2\alpha_\text{eff}^2})$.

Finally, we calculate the entanglement potential (EP) \cite{Asboth:05}. It is based on the observation that if a single-mode light field impinges on a 50:50 beam splitter together with vacuum on the other input port, then the two output ports emerge in separable states if and only if the input state is a classical state, i.e., a coherent state or a mixture of coherent states \cite{Asboth:05}. Hence, given the density matrix $\rho$ of the input state, one can calculate the density matrix $\sigma$ after the hypothetical 50:50 beam splitter and then apply any entanglement measure to $\sigma$ to quantify to which degree the input state $\rho$ is nonclassical. The EP uses the logarithmic negativity \cite{Vidal:02} as an entanglement measure. Hence, $\text{EP}= \log_2(1-2\sum_i \nu_i)$, where the $\nu_i$ are the negative eigenvalues of the partial transpose of $\sigma$. The EP is nonnegative and has no general upper bound \cite{Asboth:05}. For the CSS of Eq.\ \eqref{rho-CSS}, however, its range is $0\leq \text{EP}\leq 1$ \cite{supplemental:model:experiment}. A nonzero EP shows that the state is nonclassical. Results for the odd cat states of Fig.\ \ref{fig-Wigner} are shown as points in Fig.\ \ref{fig-entangled}(f). The largest loss-corrected EP is 0.68(2). Lines show the expectation \cite{supplemental:model:experiment}.

\textit{Limitations} --
As seen from Fig.\ \ref{fig-entangled}(a), (b) the dominant limitation for the fidelity of the generated HES is loss of target photons during state generation. The best-fit value of the loss coefficient of $L_g=20.2(11)\%$ is an order of magnitude larger than the expectation of Ref.\ \cite{Hegels:26:theo}. It turns out that for the parameters of our experiment, interaction-time broadening and incomplete blockade are responsible for a good part of this increase in $L_g$ \cite{supplemental:model:experiment}. In addition, the black and gray points in Fig.\ \ref{fig-entangled}(a) deviate from unity. This might result from self-blockade of the target light. Self-blockade and interaction-time broadening could both be mitigated considerably when increasing the duration of the target pulse, which would require a longer delay fiber. Incomplete blockade could be mitigated by reducing the radius of the atomic ensemble or by reducing laser phase noise, e.g., with the techniques of Refs.\ \cite{Nazarova:08, Denecker:25}. We expect that with these measures, the cat size in this system can be increased considerably, approaching the prediction of Ref.\ \cite{Hegels:26:theo}.

\textit{Conclusions} --
This work establishes cavity Rydberg EIT as a novel system for generating HESs. We achieve a world-record size of $\alpha_g^2=2.4$ for optical HESs. This is possible because optimizing the decoherence in this platform mitigates the dominant limitation of the experiment holding the previous record \cite{Hacker:19}. Presently, the dominant limitations are interaction-time broadening, self blockade, incomplete blockade, and laser-phase noise. We discussed ways how to overcome these limitations in the future.

\textit{Acknowledgments} --
We thank Maximilian Winter for assistance during an early phase of the experiment and Olivier Morin, Thomas Stolz, and Stephan Welte for discussions. This work was supported by Deutsche Forschungsgemeinschaft under Germany's excellence strategy via Munich Center for Quantum Science and Technology EXC-2111-390814868.

\clearpage
\onecolumngrid
\begin{center}
\large\bf
Supplemental Material: \\ Optical Schrödinger cat states generated using cavity Rydberg quantum photonics
\bigskip\medskip
\end{center}
\twocolumngrid

\renewcommand{\theequation}{S\arabic{equation}}
\renewcommand{\thefigure}{S\arabic{figure}}
\renewcommand{\thetable}{S\,\Roman{table}}
\setcounter{equation}{0}
\setcounter{figure}{0}
\setcounter{table}{0}

\section{Tomography of the CSS}

\subsection{Homodyne detection}

\label{sec-homodyne}

\begin{figure}[!t]
\centering
\includegraphics[width=\columnwidth]{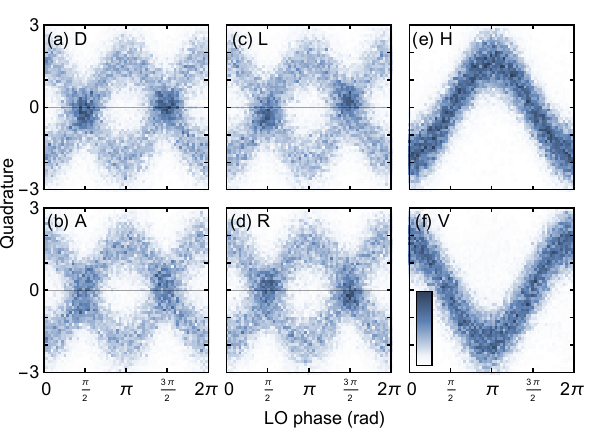}
\caption{Histograms representing sinograms for $\alpha_g^2= 1.9$ for postselection upon $D$, $A$, $L$, $R$, $H$, and $V$ in panels (a)--(f). A darker color indicates more events per bin. Two sinusoids with amplitudes $\pm\sqrt2 \alpha_d= \pm 1.8$ corresponding to $\ket{\alpha_H}$ and $\ket{\alpha_V}$ are clearly visible. In (a)--(d), the sinusoids overlap for an LO phase near $\frac\pi2$. Here, interference causes fringes with a wave vector pointing to the top of the page. For the parameters here, the fringe period is not much shorter than the width of the fringe envelope. Hence, one sees only one maximum, with a position at $q=0$,  $q<0$, and $q>0$ for $D$, $L$, and $R$, respectively. For an LO phase near $\frac{3\pi}2$, the situation is reversed. For $A$, one barely sees a minimum at $q=0$ for an LO phase near $\frac\pi2$ or $\frac{3\pi}2$. These interference effects show that the visibility $V$ is nonzero.}
\label{fig-sinogram}
\end{figure}

We use the standard scheme of a homodyne detector with a 50:50 beam splitter and two photodiodes. An electronic circuit amplifies the difference of the two photocurrents. The result is recorded using a digital oscilloscope. The photodiodes and the electronic amplification circuit were previously used in Refs.\ \cite{Hacker:19S, Hacker:phdS, Daiss:phdS} and are described there in detail, including the methods for the characterization of electronic and classical noise. The difference of the homodyne photocurrents is multiplied by a temporal mode function $u_h(t)$, described in Sec.\ \ref{sec-interaction-time}, and the result is time integrated to obtain one quadrature value \cite{Leonhardt:97S}. The collection of quadrature values recorded for many values of the local oscillator (LO) phase is referred to as a sinogram. Data are postselected upon the outcome $j\in\{D, \lb A,\lb L,\lb R,\lb H,\lb V\}$ of a measurement on the DV. Examples of histograms representing sinograms are shown in Fig.\ \ref{fig-sinogram}.

The effective loss coefficients in the detection setup are summarized in Table \ref{tab-homodyne-loss}. The optical loss includes all loss after the light left the cavity. Some part of the optical loss arises from a Faraday isolator with a transmission of 97\%. Without this Faraday isolator, a small fraction of the LO light which is unintentionally backscattered from the homodyne setup would create photon recoil heating for the atomic ensemble.

\begin{table}[!b]
\centering
\caption{Effective loss coefficients in the detection setup}\label{tab-homodyne-loss}
\begin{tabular*}{\columnwidth}{l@{\extracolsep\fill}r}
\hline \hline
Optical loss & 9 \% \\
Mode matching with local oscillator & 9 \% \\
Photodiode efficiency  & 1.5 \% \\
Electronic and classical noise & 2.9 \% \\ 
\hline
Total & $L_d=21$ \% \\
\hline \hline
\end{tabular*}
\end{table}

To characterize the transverse mode overlap with the LO, we detune the frequency of the LO from the signal light by \unit{0.2}{kHz}, send continuous-wave laser light as signal light, and block one of the photodiodes. In this way, we obtain a beat signal on the other photodiode. If the intensities of the LO and the signal light on the latter photodiode are identical and the photocurrent is not electronically high-pass filtered, then the beat visibility $V_b= \frac{I_\text{max}-I_\text{min}} {I_\text{max}+I_\text{min}}$ equals $V_b= \abs{\ip{t_\text{LO}}{t_s}} = \sqrt{F_t}$, where $\ket{t_\text{LO}}$ ($\ket{t_s}$) represent the normalized transverse modes of the LO (signal light) and $F_t$ is the fidelity between them. The corresponding loss coefficient is $L= 1-V_b^2$. We measure $V_b=95.5\%$ and obtain $L= 9\%$. The total effective loss coefficient of the detection setup $L_d$ is calculated by combining the various individual effective loss coefficients along the lines of Eq.\ \eqref{L-t}.

\subsection{Reconstruction of postselected density matrices}

\label{sec-reconstruct-CSS}

\begin{figure}[tb!]
\centering
\includegraphics[width=\columnwidth]{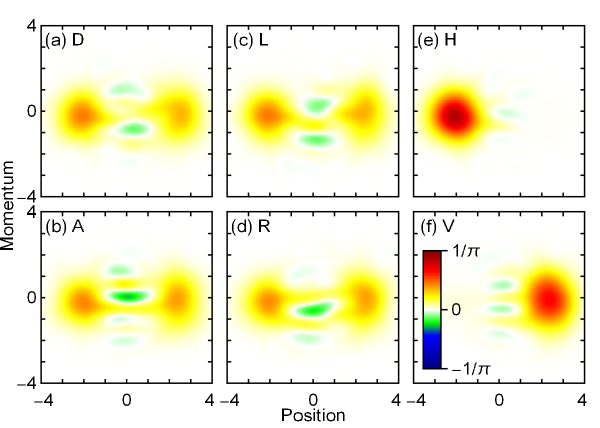}
\caption{Loss corrected Wigner function for $\alpha_g^2= 2.4$ for postselection upon $D$, $A$, $L$, $R$, $H$, and $V$ in panels (a)--(f). Peaks at $(x,p)\approx (\pm 1.9,0)$ correspond to the coherent states $\ket{\alpha_V}$ and $\ket{\alpha_H}$. Interference fringes between these peaks in panels (a)--(d) testify to the coherent nature of the superposition of these coherent states, with different panels exhibiting different phase offsets $\theta$ of the interference fringes.}
\label{fig-Wigner-CSS}
\end{figure}

The composite quantum system under consideration is a tensor product of subsystem $\mathcal A$ consisting of the DV and subsystem $\mathcal B$ consisting of the continuous variable (CV). The $R\rho R$ algorithm is used to obtain an estimator for the density matrix $\rho_j^\mathcal B$ of subsystem $\mathcal B$ postselected upon the DV-measurement outcome $j$. Examples of Wigner functions corresponding to the $\rho_j^\mathcal B$ are shown in Fig.\ \ref{fig-Wigner-CSS}

In principle, one could use filtered back projection \cite{Leonhardt:97S} instead of the $R\rho R$ algorithm to reconstruct the Wigner function from the sinogram. The information content of the Wigner function is equivalent to that of $\rho_j^\mathcal B$. Filtered back projection inherently requires low-pass filtering in phase space to reduce reconstruction noise, which would diverge without filtering. However, for reconstructing a CSS, the low-pass filtering results in an unrealistic decrease of the amplitude $V$ of the interference term, in particular for large $\alpha_\text{eff}$ \cite{Hacker:phdS}. Hence, we prefer the $R\rho R$ algorithm. The latter requires truncating Fock space and this creates some spurious interference fringes in the Wigner function, but when truncating at large enough photon number $n$, this artifact becomes small. We choose to include $0\leq n\leq 15$. This leaves us with $16^2-1=255$ independent real parameters to be determined using the $R\rho R$ algorithm.

As a crosscheck, we consider two strategies for obtaining estimators for the parameters ($\alpha_H, \alpha_V, V, \theta)$ of the CSS \eqref{rho-CSS}. One strategy consists in inferring a density matrix $\rho_j^\mathcal B$ using the $R\rho R$ algorithm followed by fitting the CSS \eqref{rho-CSS} to this density matrix. The other strategy consists in maximum-likelihood (ML) estimation of the parameters ($\alpha_H, \alpha_V, V, \theta)$ directly from the measured sinogram. 

For the latter strategy, we need the probability distribution $\Pr_{\gamma_L}(q)$ of obtaining the value $q$ in a quadrature measurement, given the CSS \eqref{rho-CSS} and the LO phase $\gamma_L$. The quadrature operator \cite{Leonhardt:97S} $\hat q_{\gamma_L} = \hat x\cos(\gamma_L)+\hat p \sin(\gamma_L)$ is a linear combination of the dimensionless position operator $\hat x= (\hat a^\dag+\hat a)/\sqrt2$ and the dimensionless momentum operator $\hat p= i(\hat a^\dag-\hat a)/\sqrt2$, where $\hat a$ is the photon annihilation operator. The eigenvalues are denoted as $q_{\gamma_L}$, $x$, and $p$, the corresponding eigenstates as $\ket{q_{\gamma_L}}$, $\ket x$, and $\ket p$, respectively. Using \cite{Leonhardt:97S}
\begin{align}
\label{ip-x-alpha}
\ip*{x}{\alpha}
= \pi^{-1/4} e^{-(x-x_0)^2/2 + ip_0 x -i p_0x_0/2}
\end{align}
with $x_0= \sqrt2 \Re(\alpha)$ and $p_0= \sqrt2 \Im(\alpha)$ and using the phase-shifting operator \cite{Leonhardt:97S}, we find
$\ip*{q_{\gamma_L}}{\alpha} = \pi^{-1/4} \lb e^{-(q_{\gamma_L}-q_0)^2/2 + i\widetilde q_0 q_{\gamma_L} -i \widetilde q_0 q_0/2}$ with
$q_0= \sqrt2 \lb \Re(\alpha e^{-i\gamma_L})$ and $\widetilde q_0= \sqrt2 \Im(\alpha e^{-i\gamma_L})$. Hence, for $\Pr_{\gamma_L}(q) = \bra{q_{\gamma_L}}\lb \rho_\theta\ket{q_{\gamma_L}}$ with $\rho_\theta$ of Eq.\ \eqref{rho-CSS}, we find
\begin{multline}
\label{Pr-q-gamma-L-alpha-V-theta}
\Pr\null_{\gamma_L}(q)
= \frac{e^{-(q-q_H)^2} + e^{-(q-q_V)^2} }{4\sqrt\pi \tau_\theta}
\\
+ \frac{V}{2\sqrt\pi \tau_\theta} e^{-(q-q_H)^2/2-(q-q_V)^2/2} 
\\  \times
\cos[(\widetilde q_H-\widetilde q_V) q +\theta + (\widetilde q_V q_V-\widetilde q_H q_H)/2]
\end{multline}
with $q_{H/V}= \sqrt2 \Re(\alpha_{H/V} e^{-i\gamma_L})$ and with $\widetilde q_{H/V} = \sqrt2 \lb \Im( \alpha_{H/V} e^{-i\gamma_L})$.

The results for the parameters ($\alpha_H, \alpha_V, V, \theta)$ obtained with the two strategies agree well with one another. As an example, we consider $\Delta V= V_\text{ML}-V$, where $V_\text{ML}$ ($V$) is the visibility obtained using direct ML estimation (the $R\rho R$ algorithm followed by fitting). Evaluating the four polarizations $D$, $A$, $L$, and $R$ for each of the HESs with six different cat sizes without loss correction, we obtain a total of 24 values of $\Delta V$. These 24 values of $\Delta V$ have a standard deviation of 0.02 and a much smaller mean value. This standard deviation is much smaller than the typical value of $V$ in Fig.\ \ref{fig-entangled}(b).

\section{Tomography of the HES}

\subsection{Generalized Stokes parameters}

The composite quantum system under consideration is a tensor product of two subsystems $\mathcal A$ and $\mathcal B$, which represent the DV and the CV, respectively. It is straightforward to show that the density matrix $\rho$ of the composite system can be expanded as, see also, e.g, Refs.\ \cite{Hacker:19S, Vlastakis:15S}
\begin{align}
\label{rho-S}
\rho
= \frac12 \sum_{i=0}^3 \sigma_i \otimes S_i
,\end{align}
where the $\sigma_i$ denote the Pauli matrices in the $(\ket H,\ket V)$ basis with $\sigma_0=\bm 1$ and where the CV operators $S_i$ are unique. The $S_i$ are Hermitian and can be regarded as a generalization of the Stokes parameters. $S_0$ is positive semidefinite and normalized to $\tr(S_0)= 1$. For a given $\rho$, one can calculate the $S_i$ as
\begin{align}
S_i
= \tr_{\mathcal A}[(\sigma_i\otimes\bm 1)\rho]
,\end{align}
where $\tr_{\mathcal A}$ denotes the partial trace over subsystem $\mathcal A$.

We consider the subensemble obtained by postselection upon the DV-measurement result $j\in\{D,\lb A,\lb L,\lb R,\lb H,\lb V\}$. This subensemble is described by the density matrix $\rho_j^\mathcal B= \frac1{\tau_j}\ev{\rho}{j}$ of the CV, where $\tau_j = \tr(\ev{\rho}{j})$ is the probability of finding the DV-measurement result $j$. Combining this with $\sigma_{0/1}= \dyad D \pm \dyad A$ and analogous expression for $\sigma_2$ and $\sigma_3$, we find
\begin{subequations}
\begin{gather}
\label{S-0-tau-rho}
S_0
= \tau_D \rho_D^\mathcal B + \tau_A \rho_A^\mathcal B
= \tau_L \rho_L^\mathcal B + \tau_R \rho_R^\mathcal B
= \tau_H \rho_H^\mathcal B + \tau_V \rho_V^\mathcal B
,\\
\label{S-1-tau-rho}
S_1
= \tau_D \rho_D^\mathcal B - \tau_A \rho_A^\mathcal B
,\\
S_2
= \tau_L \rho_L^\mathcal B - \tau_R \rho_R^\mathcal B
,\\
\label{S-3-tau-rho}
S_3
= \tau_H \rho_H^\mathcal B - \tau_V \rho_V^\mathcal B
.\end{gather}
\end{subequations}
As an aside, for the CSS of Eq.\ \eqref{rho-CSS}, we find analogously $\tau_\theta= \tr(\ev{\rho}{\psi_\theta})$ so that
\begin{align}
\label{tau-theta}
\tau_\theta
= \frac{1+V e^{-2\alpha_\text{eff}^2}\cos[\theta +\Im(\alpha_V^*\alpha_H)]}2
,\end{align}
where we used
\begin{align}
\label{ip-alpha-V-alpha-H}
\ip*{\alpha_V}{\alpha_H}
= e^{-2\alpha_\text{eff}^2+i\Im(\alpha_V^*\alpha_H)}
.\end{align}

\subsection{Reconstruction of the HES}

From the measurements, we obtain the $\rho_j^\mathcal B$ using the $R\rho R$ algorithm, as described in Sec.\ \ref{sec-reconstruct-CSS}. In addition, the measurements give us the absolute frequencies $N_j$ with which the DV-measurement outcome $j$ occurs. Hence, we obtain six obvious estimators $\tau_D= N_D/(N_D+N_A)$ etc. Inserting these into Eqs.\ \eqref{S-1-tau-rho}--\eqref{S-3-tau-rho} we obtain the estimators
\begin{subequations}
\begin{gather}
\label{S-1-estimator}
S_1
= \frac{N_D \rho_D^\mathcal B - N_A \rho_A^\mathcal B}{N_D+N_A}
,\\
S_2
= \frac{N_L \rho_L^\mathcal B - N_R \rho_R^\mathcal B}{N_L+N_R}
,\\
S_3
= \frac{N_H \rho_H^\mathcal B - N_V \rho_V^\mathcal B}{N_H+N_V}
.\end{gather}
\end{subequations}
We obtain three corresponding estimators for $S_0$, namely $\frac{1}{N_D+N_A}(N_D \rho_D^\mathcal B + N_A \rho_A^\mathcal B)$ etc. Aiming at a small statistical uncertainty, it is, of course, advantageous to give equal statistical weight to all measurement outcomes. This is achieved using the estimator
\begin{align}
\label{S-0-estimator}
S_0
= \frac{\sum_j N_j\rho_j^\mathcal B}{\sum_j N_j}
,\end{align}
which is a weighted average of the above three estimators for $S_0$ with weights $(N_D+N_A)/\sum_j N_j$ etc., where $\sum_j$ extends over $j\in\{D,A,L,R,H,V\}$.

\begin{figure}[tb!]
\centering
\includegraphics[width=\columnwidth]{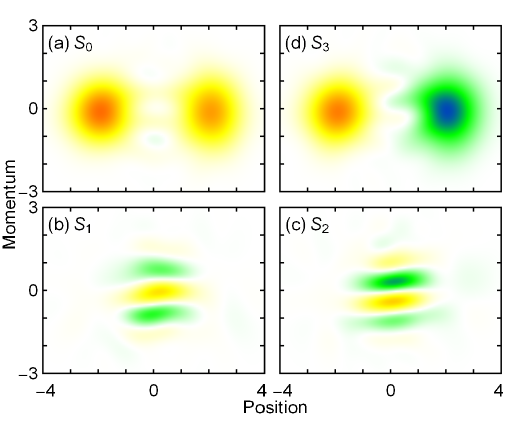}
\caption{Wigner functions corresponding to the loss-cor\-rected generalized Stokes parameters $S_0$, $S_1$, $S_2$, and $S_3$ for $\alpha_g^2=1.9$ in panels (a)--(d). The color scale is identical to Fig.\ \ref{fig-Wigner-CSS}. The two Gaussian peaks in $S_0$ represent population in the two coherent states. The fact that they correlate with $H$ and $V$ is represented by $S_3$. The interference fringes between the peaks in the Wigner functions for $D$ and $A$ contribute to $S_1$ and the $\pi/2$ phase shifted fringes in the Wigner functions for $L$ and $R$ contribute to $S_2$.}
\label{fig-Stokes}
\end{figure}

Along with $S_i$ for each $i\in\{0,1,2,3\}$, one can define a Wigner function
\begin{align}
W_i(x,p)
= \frac1{2\pi} \int_{-\infty}^\infty \mel*{x-\tfrac y2}{S_i}{x+\tfrac y2} e^{ipy} dy
\end{align}
in analog to Eq.\ \eqref{Wigner-def}. As the $S_i$ are Hermitian, the $W_i$ are real. Using Eqs.\ \eqref{S-0-tau-rho}--\eqref{S-3-tau-rho}, the triangle inequality, and that $|W(x,p)|\leq \frac1\pi$ for any density matrix \cite{Leonhardt:97S}, one finds $-\frac1\pi\leq W_i(x,p)\leq\frac1\pi$ for all $i\in\{0,1,2,3\}$. Experimental results for the $W_i$ are shown in Fig.\ \ref{fig-Stokes}.

The $R\rho R$ algorithm is designed to output a result $\rho_j^\mathcal B$, which is always positive semidefinite. However, combining six such results for different $j$ to reconstruct the HES using Eqs.\ \eqref{rho-S} and \eqref{S-1-estimator}--\eqref{S-0-estimator} is not guaranteed to produce a positive semidefinite density matrix $\rho$ of the composite system. The sum of negative eigenvalues of $\rho$ is a much-used measure of how close $\rho$ is to the nearest valid density matrix. From the data in this work, we obtain 12 density matrices of HESs, namely for six different cat sizes, each with and without loss correction. Averaging the sum of negative eigenvalues of $\rho$ over these 12 states, we obtain $-0.021$. The absolute value thereof is much smaller than 1, which indicates already that this is not a major concern here. To investigate this more thoroughly, we use renormalized eigenvalue clipping as a crude method to obtain a valid density matrix, i.e., we replace $\rho$ by $\rho_\text{pos} =(\sum_i \kappa_i \dyad*{\tilde w_i})/\sum_i \kappa_i$, where the $\kappa_i$ are the positive eigenvalues of $\rho$ and the $\ket*{\tilde w_i}$ are the corresponding eigenvectors. As a typical example, we consider the loss-corrected HES with $\alpha_g^2 = 0.38$. Here, the sum of negative eigenvalues is $-0.016$ and we find $F_\text{Bell}=82.1\%$ instead of 83.4\% and $\mathcal N=0.334$ instead of 0.341. Overall, we find that the typical corrections are small. Rather than developing a refined correction method, we choose to ignore this detail throughout this work.

\subsection{Error bars from parametric bootstrap}

We use a parametric bootstrap method to generate error bars for various quantities shown in Fig.\ \ref{fig-entangled}. We exemplify this method for the negativity $\mathcal N$. After obtaining the $N_j$ and the matrix elements $\mel{m}{\rho_j^\mathcal B}{n}$ with respect to the Fock basis from the $R\rho R$ algorithm, we pretend that the estimators for these matrix elements would give an exact description of the system. Hence, if we repeated the measurement, we would expect to find the quadrature value $q$ for a given LO phase $\gamma_L$ with probability
\begin{align}
\label{Pr-q-gamma-L-m-rho-n}
{\textstyle\Pr_{\gamma_L}^{(j)}(q)}
= \ev{\rho_j^\mathcal B}{q_{\gamma_L}}
.\end{align}
This can be calculated using $\Pr_{\gamma_L}^{(j)}(q)= \sum_{m,n=0}^\infty \lb \ip{q_{\gamma_L}}{m} \lb \mel{m}{\rho_j^\mathcal B}{n} \lb \ip{n}{q_{\gamma_L}}$ and
\begin{align}
\label{ip-n-q-LO}
\ip{q_{\gamma_L}}{n}
= e^{-i\gamma_Ln} \frac{H_n(q_{\gamma_L})}{\pi^{1/4}\sqrt{2^n n!}}\texp(-q_{\gamma_L}^2/2)
\end{align}
with the Hermite polynomial $H_n$. Equation \eqref{ip-n-q-LO} follows from \cite{Leonhardt:97S} $\ip{x}{n} = \frac{H_n(x)}{\pi^{1/4}\sqrt{2^n n!}}\texp(-x^2/2)$ together with the phase-shifting operator \cite{Leonhardt:97S}.

We use a Monte-Carlo method to obtain a sinogram with $N_j$ synthetic data points. To this end, we draw $N_j$ equally distributed values of $\gamma_L \in \null]{-}\pi,\pi]$. Along with each of them, we draw a value of $q_{\gamma_L}$ with the probability distribution of Eq.\ \eqref{Pr-q-gamma-L-m-rho-n}. We repeat this procedure for the six values of $j$. From these six synthetic sinograms, we infer six $\hat \rho_j^{\mathcal B}$ using the $R\rho R$ algorithm and from these we reconstruct $\hat \rho$. From $\hat \rho$ we calculate the quantity of interest, here $\mathcal N$, obtaining a value $\hat{\mathcal N}$. We repeat this procedure 100 times and obtain one value of $\hat{\mathcal N}$ each time. We quote the standard deviation of these 100 values of $\hat{\mathcal N}$ as the error bar on $\mathcal N$. Note that the mean values that we quote are \emph{not} generated by the parametric bootstrap method.

In the experiment, we typically take $N_j\approx 4\times10^4$ data points for each $j$ for each cat size. As the error bars in Fig.\ \ref{fig-entangled} are much smaller than the mean values, we conclude that $N_j$ is large enough to avoid problems that would arise for too small $N_j$.

\section{Expectations for Fig.\ \ref{fig-entangled}}

\subsection{Bell-state fidelity}

\label{sec-Bell}

Here, we discuss how we choose the Bell state for calculating the Bell-state fidelity in Fig.\ \ref{fig-entangled}(a). We choose the orthonormal sequence $(\ket\up,\ket\dn)$ for given best-fit values $\alpha_H$ and $\alpha_V$ such that the fidelity between the Bell state 
\begin{align}
\label{psi-B}
\ket{\psi_B}
= \frac{\ket{H,\up}+\ket{V,\dn}}{\sqrt2}
\end{align}
and the HES \eqref{rho-HES} is maximized. Compared to maximizing the fidelity between the Bell state and the \emph{measured} HES, this strategy has the advantage that the optimization can be carried out analytically. 

Before searching for $(\ket\up,\ket\dn)$, we first show that an arbitrary Bell state can be written in the form of Eq.\ \eqref{psi-B}. By definition, an arbitrary Bell state can be written as $\frac1{\sqrt2}(\ket{0_1,0_2}+\ket{1_1,1_2})$, where $(\ket{0_1},\ket{1_1})$ and $(\ket {0_2},\ket {1_2})$ are arbitrary orthonormal sequences. Obviously, there must be a unitary transformation $U\in\operatorname{U}(2)$ such that $\smqty(\ket{0_1}\\\ket{1_1}) = U \smqty(\ket H\\ \ket V)$. It is easy to show that this arbitrary Bell state can be rewritten as Eq.\ \eqref{psi-B} with $\smqty(\ket{\up}\\\ket{\dn}) = U^T \smqty(\ket {0_2}\\ \ket {1_2})$. As $U$ is unitary, so is its transpose $U^T$. As $(\ket {0_2},\ket {1_2})$ is an arbitrary orthonormal sequence, so is $(\ket {\up},\ket {\dn})$. Hence, a state is an arbitrary Bell state if and only if it can be written in the form of Eq.\ \eqref{psi-B} with an arbitrary orthonormal sequence $(\ket {\up},\ket {\dn})$. Hence, it suffices to consider Bell states of this form.

$F_\text{Bell} = \ev{\rho}{\psi_B}$ with $\rho$ of Eq.\ \eqref{rho-HES} reads
\begin{align}
\label{F-B}
F_\text{Bell}
= \frac{\abs{\ip{\up}{\alpha_H}}^2 + \abs{\ip{\dn}{\alpha_V}}^2}4 + V \frac{\ip{\up}{\alpha_H}\ip{\alpha_V}{\dn} + \cc}4
.\end{align}
We search the orthonormal sequence $(\ket\up,\ket\dn)$ which maximizes $F_\text{Bell}$ for given $(\ket{\alpha_H},\ket{\alpha_V})$ with $\abs{\ip{\alpha_H}{\alpha_V}}\in\null]0,1[$. Obviously, $\ket{\up}$ and $\ket{\dn}$ must be elements of the 2D subspace (of Fock space) spanned by $(\ket{\alpha_H},\ket{\alpha_V})$. We restrict the considerations to this 2D subspace.

We expand the vectors $\ket{\alpha_H}$ and $\ket{\alpha_V}$ in the -- so far unknown -- orthonormal basis $(\ket{\up} , \ket{\dn})$ of the 2D subspace. As the vectors $\ket{\alpha_H}$ and $\ket{\alpha_V}$ are normalized, we can parameterize this expansion as
\begin{align}
\label{alpha-HV-vs-updn}
\mqty(\ket{\alpha_H} \\ \ket{\alpha_V})
= \mqty(e^{i\varphi_{11}}c_1 & e^{i\varphi_{12}}s_1 \\ e^{i\varphi_{21}}s_2 & e^{i\varphi_{22}}c_2)
\mqty(\ket{\up} \\ \ket{\dn})
\end{align}
with $c_{1/2}= \cos\theta_{1/2}$ and $s_{1/2}= \sin\theta_{1/2}$ with $\theta_1,\theta_2\in [0,\frac\pi2]$ and with $\varphi_{11},\varphi_{12},\varphi_{21},\varphi_{22}\in\mathbb R$. For the moment, we restrict the considerations to $\theta_1,\theta_2\in \null]0,\frac\pi2[$. We will turn to the domain boundary at the very end of this discussion. In addition, we exclude the case $V=0$ and the case $V=1$, because they are of little experimental relevance. Without loss of generality, we choose $\varphi_{11}= 0$ by resetting the irrelevant global phase of $\ket{\psi_B}$. We abbreviate
\begin{align}
\varphi
= \arg (\ip{\alpha_H}{\alpha_V})
.\end{align}
Inserting Eq.\ \eqref{alpha-HV-vs-updn} into the lefthand side of $\ip{\alpha_H}{\alpha_V}= e^{i\varphi} \abs{\ip{\alpha_H}{\alpha_V}}$, we obtain the constraint $g_0=0$ with $g_0= e^{i\varphi_{21}-i\varphi} c_1 s_2 + e^{i\varphi_{22}-i\varphi_{12}-i\varphi} s_1 c_2- \abs{\ip{\alpha_H}{\alpha_V}}$. We want to maximize $F_\text{Bell}= \frac14(c_1^2 + c_2^2+2V c_1c_2\cos\varphi_{22})$ subject to this constraint.

We rewrite the constraint as two real-valued constraints $g_1=g_2=0$ with $g_1= \Re(g_0)$ and $g_2= \Im(g_0)$. Using the method of Lagrange multiplies, we find that, if $F_\text{Bell}$ has an extremum, then $\exists \lambda_1,\lambda_2\in\mathbb R$ such that $\lambda_1 \nabla g_1 + \lambda_2 \nabla g_2= \nabla F_\text{Bell}$ where $\nabla = (\partial_{\theta_1}, \lb \partial_{\theta_2}, \lb \partial_{\varphi_{12}},\lb \partial_{\varphi_{21}}, \lb \partial_{\varphi_{22}})^T$. Obviously, $\partial_{\varphi_{22}} g_0= -\partial_{\varphi_{12}}g_0$. Hence $\partial_{\varphi_{22}} F_\text{Bell} = -\partial_{\varphi_{12}} F_\text{Bell}$. This reads $-\frac12 V c_1c_2 \sin\varphi_{22}=0$. Hence $\sin\varphi_{22}=0$. We abbreviate $\sigma_1= \cos\varphi_{22}$ and find $\sigma_1\in\{-1,1\}$. The derivatives $\partial_{\theta_1}$ and $\partial_{\theta_2}$ give $G_\theta \smqty(\lambda_1 \\ \lambda_2) = F_\theta \smqty(c_1 \\ c_2)$ with $F_\theta= \frac12 \smqty(1 & \sigma_1 V \\ \sigma_1 V & 1)$ and with some $G_\theta\in\mathbb R^{2\times2}$. Using $(c_1,c_2)\neq(0,0)$ and $\det(F_\theta)\neq0$, we find $(\lambda_1, \lambda_2)\neq(0,0)$. The derivatives $\partial_{\varphi_{12}}$ and $\partial_{\varphi_{21}}$ give $G_\varphi \smqty(\lambda_1 \\ \lambda_2) = \smqty(0 \\ 0)$ with $G_\varphi = \smqty(s_1c_2 \sin(\varphi_{22}-\varphi_{12}-\varphi) & -s_1c_2 \cos(\varphi_{22}-\varphi_{12}-\varphi) \\ -c_1s_2 \sin(\varphi_{21}-\varphi) & c_1s_2 \cos(\varphi_{21}-\varphi))$. Using $(\lambda_1, \lambda_2)\neq(0,0)$, we find $\det(G_\varphi) =0$. Hence $\sin(\varphi_{22}- \varphi_{12} - \varphi_{21})=0$. We abbreviate $\sigma_2= \cosine(\varphi_{22}- \varphi_{12} - \varphi_{21})$ and find $\sigma_2 \in\{-1,1\}$. The constraint $g_0=0$ becomes $e^{i\varphi -i\varphi_{21}} \lb \abs{\ip{\alpha_H}{\alpha_V}} = \sin(\sigma_2\theta_1+\theta_2)$. Hence $e^{i\varphi -i\varphi_{21}}$ is real. We abbreviate $\sigma_3= e^{i\varphi -i\varphi_{21}}$ and find $\sigma_3\in\{-1,1\}$. We abbreviate $\sigma_4= \sigma_1 \sigma_2 \sigma_3$. We conclude that, if $F_\text{Bell}$ has an extremum, then $\exists \sigma_1,\sigma_3,\sigma_4\in\{-1,1\}$ such that
\begin{align}
\mqty(\ket{\alpha_H} \\ \ket{\alpha_V})
= \mqty(c_1 & \sigma_4 s_1 e^{-i\varphi} \\ \sigma_3 s_2 e^{i\varphi} & \sigma_1 c_2 )
\mqty(\ket{\up} \\ \ket{\dn})
.\end{align}
We absorb $\sigma_4,\sigma_3$ in the signs of $s_1,s_2$ by extending the domain of $\theta_1,\theta_2$ to $]{-}\frac\pi2,0[ \null \cup \null ]0,\frac\pi2[$. We obtain
\begin{align}
\label{alpha-HV-vs-updn-real}
\mqty(\ket{\alpha_H} \\ e^{-i\varphi}\ket{\alpha_V})
= \mqty(c_1 & s_1 \\ s_2 & \sigma_1 c_2)
\mqty(\ket{\up} \\ e^{-i\varphi} \ket{\dn})
.\end{align}
Note that $(\ket{\up} , e^{-i\varphi} \ket{\dn})$ is an orthonormal basis. Now, the constraint reads $g_0 =0$ with $g_0 = g_1= \sine(\sigma_1\theta_1+\theta_2)-\abs{\ip{\alpha_H}{\alpha_V}}$, while $g_2$ vanishes automatically. As an aside, now all elements of the $2\times2$ matrix are real, which means that the remaining problem is equivalent to an optimization with $\ket \up$, $e^{-i\varphi}\ket\dn$, $\ket {\alpha_H}$, and $e^{-i\varphi}\ket{\alpha_V}$ replaced by vectors in $\mathbb R^2$.

Returning to the above Lagrange multipliers, we find $\lambda_1(\partial_{\theta_1}-\sigma_1\partial_{\theta_2})g_1 = (\partial_{\theta_1}- \sigma_1\partial_{\theta_2})F_\text{Bell}$, because $g_2$ vanishes automatically. This reads $0= \frac12(1-V)(c_1-\sigma_1c_2)$. This gives $c_1= \sigma_1 c_2$. Using the domain of $\theta_1,\theta_2$, we find that $c_1= \sigma_1 c_2$ is equivalent to $\sigma_1=1$ and $\theta_2= \sigma_5\theta_1$ with $\sigma_5\in\{-1,1\}$. Hence, at the extremum $F_\text{Bell} = \frac{1+V}{2} c_1^2$. For $\sigma_5=-1$, the constraint would become $\abs{\ip{\alpha_H}{\alpha_V}}=0$, which is in conflict with $\abs{\ip{\alpha_H}{\alpha_V}}\in\null]0,1[$. Hence, $\sigma_5=1$ and the constraint becomes $\sin(2\theta_1)=\abs{\ip{\alpha_H}{\alpha_V}}$. Given the domain of $\theta_1$, this is equivalent to $\theta_1= \frac\zeta2$ or $\theta_1= \frac{\pi-\zeta}2$, where we abbreviated
\begin{align}
\zeta
= \arcsin \abs{\ip{\alpha_H}{\alpha_V}}
,&&
c
= \cos\frac\zeta2
,&&
s
= \sin\frac\zeta2
.\end{align}
Hence $\zeta\in\null]0,\frac\pi2[$. Using $c^2-s^2=\cos\zeta>0$, we find that $F_\text{Bell}|_{\theta_1=\frac\zeta2}= \frac{1+V}{2} c^2$ is larger than $F_\text{Bell}|_{\theta_1=\frac{\pi-\zeta}2}= \frac{1+V}{2} s^2$. Finally, we find that $F_\text{Bell}$ at the domain boundary, $\theta_1 \in\{0,\frac\pi2\}$ or $\theta_2\in\{0,\frac\pi2\}$, is smaller than $\frac{1+V}{2} c^2$. Hence, we obtain the global maximum of $F_\text{Bell}$ for 
\begin{align}
\mqty(\ket{\alpha_H} \\ \ket{\alpha_V})
= \mqty(c & s e^{-i\varphi} \\ s e^{i\varphi} & c)
\mqty(\ket{\up} \\ \ket{\dn})
.\end{align}
To calculate $( \ket{\up}, \ket{\dn})$ from given $(\ket{\alpha_H}, \ket{\alpha_V})$, we use the inverse transformation
\begin{align}
\mqty( \ket{\up} \\ \ket{\dn})
= \frac1{c^2-s^2}\mqty( c & -se^{-i\varphi} \\ -se^{i\varphi} & c ) 
\mqty(\ket{\alpha_H} \\ \ket{\alpha_V})
.\end{align}
The maximum is $\max(F_\text{Bell}) = \frac{1+V}{2} c^2$, where $c^2= \frac12(1+\sqrt{1-\sin^2 \zeta})$ with $\sin^2 \zeta = \abs{\ip{\alpha_H}{\alpha_V}}^2 = e^{-4\alpha_\text{eff}^2}$, where we used Eq.\ \eqref{ip-alpha-V-alpha-H}. Hence
\begin{align}
\max(F_\text{Bell})
= \frac{1+V}{4} \left(1+\sqrt{1-e^{-4\alpha_\text{eff}^2}}\right)
.\end{align}

\subsection{Negativity}

We turn to the negativity $\mathcal N$. First, we discuss its maximum. We note that according to proposition 1 of Ref.\ \cite{Vidal:02S}, $\mathcal N(\rho)$ is a convex function. This implies that there exists a pure state, which reaches the maximum of $\mathcal N$. Any pure state has a Schmidt decomposition $\ket\psi = \sum_{i=1}^k c_i \ket{i_\mathcal A,i_\mathcal B}$, where $k$ is the Schmidt number, the $c_i>0$ are the Schmidt coefficients with $\sum_{i=1}^k c_i^2=1$, and the $\ket{i_\mathcal A}$ and $\ket{i_\mathcal B}$ form suitable orthonormal sequences. For a tensor product of spaces of dimensions $d_1$ and $d_2$ one finds $\max(k)= \min\{d_1,d_2\}$ because an orthonormal sequence of length $k$ can only exist in a space of dimension $d\geq k$. According to proposition 8 of Ref.\ \cite{Vidal:02S}, the negativity of this pure state is $\mathcal N= \frac12[(\sum_{i=1}^k c_i)^2-1]$. Hence, we need to vary all the $c_i\in\mathbb R^+$ to maximize $(\sum_{i=1}^k c_i)^2$ subject to the constraint $\sum_{i=1}^k c_i^2=1$. Using Lagrange multipliers, we find a maximum, if all $c_i= 1/\sqrt k$. This turns out to be the global maximum. Hence, for a given $k$, the maximum of $\mathcal N$ is $(k-1)/2$ so that
\begin{align}
\max(\mathcal N)
= \frac{\min\{d_1,d_2\}-1}2
.\end{align}

Next, we calculate the negativity of the HES of Eq.\ \eqref{rho-HES}. We take the partial transpose of the HES \eqref{rho-HES} with respect to the orthonormal basis $(\ket H,\ket V)$ of subsystem $\mathcal A$. We obtain
\begin{multline}
\label{rho-T-A}
\rho^{T_\mathcal A}
= \tfrac12 \dyad{H,\alpha_H} +\tfrac12 \dyad{V,\alpha_V}
\\
+ \tfrac12 V (\dyad{V,\alpha_H}{H,\alpha_V} + \Hc)
.\end{multline}
The linearly independent sequence $\mathcal U:(\ket{u_1}, \lb \ket{u_2}, \lb \ket{u_3}, \lb \ket{u_4})= (\ket{H,\alpha_H}, \lb \ket{H,\alpha_V}, \lb \ket{V,\alpha_H}, \lb \ket{V,\alpha_V})$ spans a subspace which is invariant under applying $\rho^{T_\mathcal A}$. From now on, we restrict $\rho^{T_\mathcal A}$ to this two-qubit subspace. Hence, $\mathcal U$ becomes a basis. Its metric tensor reads 
\begin{align}
g_{jk}^{(\mathcal U)}
= \ip{u_j}{u_k}
= \bm 1 \otimes \mqty( 1 & s_0 \\ s_0^* & 1 )
=
\mqty(1 & s_0 & 0 & 0 \\
s_0^* & 1 & 0 & 0 \\
0 & 0 & 1 & s_0 \\
0 & 0 & s_0^* & 1)
,\end{align}
where we abbreviated $s_0 = \ip{\alpha_H}{\alpha_V}$ and used the Kronecker product of matrices. Equation \eqref{rho-T-A} is of the form $\rho^{T_\mathcal A}= \sum_{i,j=1}^4 \ket{u_i} (\rho^{T_\mathcal A})^{ij} \bra{u_j}$. The matrix representation $(\rho^{T_\mathcal A})^i\null_k$ with respect to the basis $\mathcal U$ is defined by $\rho^{T_\mathcal A} \ket{u_k}= \sum_{i=1}^4 \ket{u_i} (\rho^{T_\mathcal A})^i\null_k$. Hence $(\rho^{T_\mathcal A})^i\null_k= \sum_{j=1}^4 (\rho^{T_\mathcal A})^{ij} g_{jk}^{(\mathcal U)}$ and
\begin{align}
\label{rho-T-A-i-k}
(\rho^{T_\mathcal A})^i\null_k
= \frac12\mqty(
1 & 0 & 0 & 0 \\
0 & 0 & V & 0 \\
0 & V & 0 & 0 \\
0 & 0 & 0 & 1 )
g_{jk}^{(\mathcal U)}
= 
\frac12
\mqty(
1 & s_0 & 0 & 0 \\
0 & 0 & V & s_0 V \\
s_0^* V & V & 0 & 0 \\
0 & 0 & s_0^* & 1 )
.\end{align}
While the operator $\rho^{T_\mathcal A}$ is Hermitian, this matrix representation is not. This is because the basis $\mathcal U$ is not orthonormal. This matrix representation can be diagonalized analytically. The eigenvalues are $\lambda'_{1/2}= \frac14 [1-V \pm \sqrt{(1+V)^2-4V|s_0|^2} ]$ and $\lambda'_{3/4}= \frac14 [ 1+V \pm \sqrt{(1-V)^2+4V|s_0|^2} ]$. Only $\lambda'_2$ is negative. With $|s_0|^2 = \abs{\ip{\alpha_H}{\alpha_V} }^2= e^{-4\alpha_\text{eff}^2}$ we find
\begin{align}
\mathcal N
= -\lambda'_2
= \frac{-1+V + \sqrt{(1+V)^2-4V e^{-4\alpha_\text{eff}^2}}}4
.\end{align}
For $V=1$ and $\alpha_\text{eff}\to\infty$, this approaches $\mathcal N\to \frac12$.

\subsection{Macroscopicity}

To calculate the macroscopicity of the HES of Eq.\ \eqref{rho-HES}, we decompose $\mathcal I = \mathcal I_+ - \mathcal I_-$ with $\mathcal I_+= \ttr [(\rho \hat a^\dag) (\hat a\rho)]$ and $\mathcal I_-= \ttr [\rho (\hat a\rho \hat a^\dag)]$. Using $\hat a\ket \alpha = \alpha \ket \alpha$, we find for the HES of Eq.\ \eqref{rho-HES}
\begin{subequations}
\begin{align}
\mathcal I_+
&
= \tfrac14\tr \big( |\alpha_H|^2 A_0^2 + |\alpha_V|^2 B_0^2 +  (|\alpha_H|^2  + |\alpha_V|^2) C_0^\dag C_0 \big)
,\\
\mathcal I_-
&
= \tfrac14\tr \big( |\alpha_H|^2 A_0^2 + |\alpha_V|^2 B_0^2 + (\alpha_H\alpha_V^* +\cc) C_0^\dag C_0 \big)
,\end{align}
\end{subequations}
where we abbreviated $A_0= \dyad{H,\alpha_H}$, $B_0= \dyad{V,\alpha_V}{V,\alpha_V}$, and $C_0= V \dyad{V,\alpha_V}{H,\alpha_H}$ and we used $\tr(A_0B_0)= \tr(A_0C_0)= \tr(B_0C_0)= \ttr(C_0 C_0)= 0$. Hence
\begin{align}
\mathcal I
= \alpha_\text{eff}^2 V^2
,\end{align}
where we used $\alpha_\text{eff}= |\alpha_H-\alpha_V|/2$ and $\ttr(C_0^\dag C_0)= V^2$. Obviously, $\mathcal I$ has no upper bound.

\subsection{Minimum of the Wigner function}

The Wigner function of a single-mode light field is \cite{Leonhardt:97S}
\begin{align}
\label{Wigner-def}
W(x,p)
= \frac1{2\pi} \int_{-\infty}^\infty \mel*{x-\tfrac y2}{\rho}{x+\tfrac y2} e^{ipy} dy
.\end{align}
Using Eq.\ \eqref{ip-x-alpha}, we find for the CSS of Eq.\ \eqref{rho-CSS}
\begin{multline}
\label{W-x-p}
W(x,p)
= \frac{1}{4\pi \tau_\theta}
\Big(
e^{-(x-x_H)^2-(p-p_H)^2} + e^{-(x-x_V)^2}
\\
\times  e^{-(p-p_V)^2}
+ 2 V e^{-(x-x_c)^2- (p-p_c)^2} \cos (\theta_W ) \Big)
,\end{multline}
where $x_{H/V}= \sqrt2 \Re(\alpha_{H/V})$, $p_{H/V}= \sqrt2 \Im(\alpha_{H/V})$, $x_c= (x_H+x_V)/2$, and $p_c= (p_H+p_V)/2$. In addition
\begin{align}
\label{theta-W}
\theta_W
= \theta + \Im(\alpha_V^*\alpha_H) + \bm k \cdot \mqty( x-x_c \\ p-p_c )
\end{align}
with
\begin{align}
\label{wave-vector-in-phase-space}
\bm k
= \mqty( p_H-p_V \\ x_V-x_H  )
.\end{align}
$\bm k$ can be interpreted as a wave vector in phase space. $\bm k$ is given by the locations of the two Gaussian peaks representing the two coherent states.

Obviously, the cosine in Eq.\ \eqref{W-x-p} has a $\{\smqty{\text{maximum} \\ \text{minimum}}\}$ at $(x,p)=(x_c,p_c)$ if and only if
\begin{align}
\label{even-odd-cat}
e^{i\theta+i\Im(\alpha_V^*\alpha_H)} 
= \pm 1
.\end{align}
We refer to this as the $\{\smqty{\text{even} \\ \text{odd}}\}$ cat state. Note that the phase $\theta+\Im(\alpha_V^*\alpha_H)$ of Eq.\ \eqref{even-odd-cat} appears in $\tau_\theta$ in Eq.\ \eqref{tau-theta}, too. In our experiment $\alpha_V\approx -\alpha_H$, which implies $\Im(\alpha_V^*\alpha_H)\approx 0$. Hence, the criterion for the $\{\smqty{\text{even} \\ \text{odd}}\}$ cat state reduces to $e^{i\theta}=\pm1$, as mentioned in the letter.

The term $\Im(\alpha_V^*\alpha_H)$ in Eq.\ \eqref{even-odd-cat} can be regarded as originating from the fact that the product of two displacement operators $\hat D(\alpha)$ \cite{Leonhardt:97S} contains a nontrivial phase factor. More precisely, using the Baker-Campbell-Hausdorff formula, one finds 
\begin{align}
\hat D(\alpha_1) \hat D(\alpha_2) 
= e^{-i\Im(\alpha_1^*\alpha_2)} \hat D(\alpha_1+\alpha_2)
.\end{align}
Hence, considering the pure state $\ket{\psi_\text{CSS}} = (\ket{\alpha_V}+e^{i\theta}\ket{\alpha_H})/\sqrt{4\tau_\theta}$, which is the state of Eq.\ \eqref{rho-CSS} for $V=1$, and setting $\alpha_c= \frac12(\alpha_H+\alpha_V)$ and $\alpha_0= \frac12(\alpha_H-\alpha_V)$, and using \cite{Leonhardt:97S} $\hat D(\alpha)\ket0=\ket\alpha$ for any coherent state $\ket\alpha$, we find $\hat D(-\alpha_c) \ket{\psi_\text{CSS}} = e^{-i \Im(\alpha_V^*\alpha_H)/2} (\ket{-\alpha_0}+ e^{i\theta +i \Im(\alpha_V^*\alpha_H)} \lb \ket{\alpha_0})/\sqrt{4\tau_\theta}$, which is the state of Eq.\ \eqref{rho-CSS} with $\theta$ replaced by $\theta+\Im(\alpha_V^*\alpha_H)$.

For the odd cat state of Eq.\ \eqref{even-odd-cat}, we find from Eq.\ \eqref{W-x-p}, see also Ref.\ \cite{Hacker:phdS}
\begin{align}
W(x_c,p_c)
= \frac{e^{-2\alpha_\text{eff}^2}-V}{2\pi\tau_\theta}
= \frac{e^{-2\alpha_\text{eff}^2}-V}{\pi(1-V e^{-2\alpha_\text{eff}^2})}
.\end{align}
If this is negative, this is the global minimum of $W(x,p)$. Otherwise, the infimum of $W(x,p)$ is 0 and it is approached for $x^2+p^2\to\infty$.

\subsection{Entanglement potential}

Here, we calculate the EP of the CSS of Eq.\ \eqref{rho-CSS}. The density matrix $\sigma$ of the composite system consisting of the two output modes of the hypothetical 50:50 beam splitter reads
\begin{multline}
\sigma
= \frac{\dyad{\alpha,\alpha} + \dyad{\beta,\beta} }{4\tau_\theta}
\\
+ V \frac{e^{i\theta} \dyad{\alpha,\alpha}{\beta,\beta} + \Hc}{4\tau_\theta}
\end{multline}
with coherent-state amplitudes $\alpha= \alpha_H/\sqrt2$ and $\beta = \alpha_V\sqrt2$. We take the partial transpose of $\sigma$ with respect to the orthonormal basis of Fock states $\ket n$ of subspace $\mathcal A$. This partial transpose has the property $(\ket{\alpha_1,v} \lb \bra{\alpha_2,w})^{T_\mathcal A} = \dyad{\alpha_2^*,v}{\alpha_1^*,w}$ for coherent states $\ket{\alpha_1}$, $\ket{\alpha_2}$ and arbitrary states $\ket v$, $\ket w$. Hence
\begin{multline}
\sigma^{T_\mathcal A}
= \frac{\dyad{\alpha^*,\alpha} + \dyad{\beta^*,\beta} }{4\tau_\theta}
\\
+ V \frac{e^{i\theta} \dyad{\beta^*,\alpha}{\alpha^*,\beta} + \Hc }{4\tau_\theta}
.\end{multline}
Similar to Eq.\ \eqref{rho-T-A}, the linearly independent sequence $\mathcal V:(\ket{v_1}, \lb \ket{v_2}, \lb \ket{v_3}, \lb \ket{v_4})= (\ket{\alpha^*,\alpha}, \lb \ket{\alpha^*, \beta}, \lb \ket{\beta^*,\alpha}, \lb \ket{\beta^*,\beta})$ spans a subspace which is invariant under applying $\sigma^{T_\mathcal A}$. The metric tensor of $\mathcal V$ is
\begin{align}
g_{jk}^{(\mathcal V)}
&
= \ip{v_j}{v_k}
= \mqty( 1 & s_3^* \\ s_3 & 1 ) \otimes \mqty( 1 & s_3 \\ s_3^* & 1 )
\notag \\ &
=
\mqty(1 & s_3 & s_3^* & |s_3|^2 \\
s_3^* & 1 & (s_3^*)^2 & s_3^* \\
s_3 & s_3^2 & 1 & s_3 \\
|s_3|^2 & s_3 & s_3^* & 1)
,\end{align}
where we abbreviated $s_3 = \ip{\alpha}{\beta}$ and we used $\ip{\alpha_1^*}{\alpha_2^*} = \ip{\alpha_1}{\alpha_2}^*$ for coherent states $\ket{\alpha_1}$, $\ket{\alpha_2}$. In analog to Eq.\ \eqref{rho-T-A-i-k}, we find the matrix representation
\begin{align}
(\sigma^{T_\mathcal A})^i\null_k
&
= \frac1{4\tau_\theta}
\mqty(
1 & 0 & 0 & 0 \\
0 & 0 & V_1 & 0 \\
0 & V_1^* & 0 & 0 \\
0 & 0 & 0 & 1 )
g_{jk}^{(\mathcal V)}
\notag \\ &
= 
\frac1{4\tau_\theta}
\mqty(
1 & s_3 & s_3^* & |s_3|^2 \\
V_1 s_3 & V_1s_3^2 & V_1 & V_1s_3 \\
V_1^* s_3^* & V_1^* & V_1^* (s_3^*)^2 & V_1^* s_3^* \\
|s_3|^2 & s_3 & s_3^* & 1
)
,\end{align}
where we abbreviated $V_1= Ve^{-i\theta}$. We note that $(1,\lb 0 ,\lb 0,\lb -1)^T$ is an eigenvector. We find analytic solutions for all eigenvalues if $V_1 s_3^2\in\mathbb R$ because in this case there is an eigenvector of the form $(0,c_2,c_3,0)^T$ with $c_2,c_3\in\mathbb C$. Using $s_3^2= (e^{-\frac12\abs{\alpha-\beta}^2+i\Im(\alpha^*\beta)})^2 = e^{-\frac12\abs{\alpha_H-\alpha_V}^2+i\Im(\alpha_H^*\alpha_V)} = \ip{\alpha_H}{\alpha_V}$, we find that $V_1 s_3^2\in\mathbb R$ is equivalent to $e^{i\theta+i\Im(\alpha_V^*\alpha_H)} =\pm 1$, which characterizes the $\{\smqty{\text{even} \\ \text{odd}}\}$ cat state according to Eq.\ \eqref{even-odd-cat}. Here, we are interested in the odd cat state and we find 
\begin{multline}
\operatorname{EP}_\text{odd}
= \log_2 \bigg[ \frac1{2(1-Ve^{-2\alpha_\text{eff}^2})} \bigg( (1 +V)(1 -e^{-2\alpha_\text{eff}^2}) 
\\ 
+ \sqrt{(1 +V)^2(1 + e^{-2\alpha_\text{eff}^2})^2 -16 V e^{-2\alpha_\text{eff}^2}} \bigg) \bigg]
.\end{multline}
As an aside, for the even cat state, we find
\begin{align}
\operatorname{EP}_\text{even}
= \log_2 \left( \frac{1+V}{1+Ve^{-2\alpha_\text{eff}^2}} \right)
.\end{align}
$\operatorname{EP}_\text{odd}$ and $\operatorname{EP}_\text{even}$ are each maximized for $V=1$ and $\alpha_\text{eff}\to\infty$, where they each approach $\operatorname{EP}\to 1$.

\section{Cavity Rydberg EIT}

In this section, we discuss several aspects related to cavity Rydberg EIT. In Sec.\ \ref{sec-CREIT-sub} we set the stage by giving a brief summary of the cavity Rydberg EIT model of Ref.\ \cite{Hegels:26:theoS}. In Sec.\ \ref{sec-coherence-time} we present a novel method for measuring the coherence time. In Sec.\ \ref{sec-electric-field} we use the outcome of this coherence-time measurement to derive an upper bound on the electric-field noise in the setup. In Secs.\ \ref{sec-incomplete-blockade} and \ref{sec-interaction-time} we model incomplete Rydberg blockade and interaction-time broadening, respectively, to obtain an expectation for the effective loss coefficient $L_g$. In Sec.\ \ref{sec-vary-C-Lambda} we discuss how these two effects depend on $C$ and $\Lambda_V$.

\subsection{Brief summary}

\label{sec-CREIT-sub}

We start with a brief summary of some key results of the cavity Rydberg EIT model of Ref.\ \cite{Hegels:26:theoS}. The cavity reflection coefficient in the presence/absence of Rydberg blockade is
\begin{align}
\label{R-HV}
\mathcal R_{H/V}
= -1+2\eta_\text{esc}\mu_{H/V}
\end{align}
with
\begin{align}
\mu_{H/V}
= \frac{1}{1-i\Delta_\text{cav}/\kappa+C \nu_{H/V}}
\end{align}
and
\begin{align}
\nu_{H/V}
= \frac{\gamma}{\gamma-i\Delta_s+\frac{\Omega_c^2}{2\gamma_{rg}-4i\Delta_{2,H/V}}}
.\end{align}
Here $C= N_a g^2/\kappa\gamma$ is the collective cooperativity, $N_a$ is the atom number, $2g$ is the single-atom vacuum Rabi frequency, $\Omega_c$ is the Rabi frequency of the EIT coupling light, $\Delta_\text{cav}= \omega_\text{in}-\omega_\text{cav}$ is the detuning between the EIT signal light and the cavity resonance, $\Delta_s=\omega_\text{in}-\omega_{eg}$ is the detuning between the EIT signal light and the atomic transition $\ket g \leftrightarrow \ket e$, $\Delta_c = \omega_c-\omega_{re}$ is the detuning between the EIT coupling light and the atomic transition $\ket e \leftrightarrow \ket r$, $\Delta_{2,V}= \Delta_c+\Delta_s$ is the two-photon detuning for the two-photon transition $\ket g \leftrightarrow \ket r$, $1/2\kappa$ is the $1/e$ lifetime of a resonant photon in the cavity in the absence of atoms, $1/2\gamma$ is the $1/e$ lifetime of population in the atomic state $\ket e$, $1/\gamma_{rg}$ is the coherence time for superpositions of states $\ket g$ and $\ket r$, and $\eta_\text{esc}$ is the probability that a resonant photon prepared in the cavity in the absence of atoms escapes from the one-sided cavity through the input-output coupler. Rydberg blockade causes $\Delta_{2,H}$ to differ from $\Delta_{2,V}$. For example, complete Rydberg blockade corresponds to $|\Delta_{2,H}|\to\infty$.

The amplitude of the light spontaneously emitted by the atoms is $a_{H/V}= \mathcal A_{H/V} \alpha_\text{in}$ with
\begin{align}
\label{A-HV}
\mathcal A_{H/V}
= 2 \mu_{H/V} \sqrt{\eta_\text{esc} C \Re(\nu_{H/V})} e^{i\arg(\nu_{H/V})}
.\end{align}
The amplitude of the light lost from the highly reflective mirrors is $m_{H/V}= \mathcal M_{H/V} \alpha_\text{in}$ with
\begin{align}
\label{M-HV}
\mathcal M_{H/V}
= 2\mu_{H/V} \sqrt{\eta_\text{esc} (1-\eta_\text{esc})}
.\end{align}

Our experiment is carried out in the parameter regime
\begin{align}
\label{parameter-regime}
1 
\ll C
\ll \Lambda_V^2
,&&
\gamma_{rg}
\ll \Omega_c
,&&
1-\eta_\text{esc}
\ll 1 
.\end{align}
In addition, we choose $\omega_\text{cav}=\omega_{eg}$ and $\Delta_c=0$ so that
\begin{align}
\label{Delta-all-identical}
\Delta_s
= \Delta_\text{cav}
= \Delta_{2,V}
.\end{align}
In the following, we always assume that Eq.\ \eqref{Delta-all-identical} holds. Together with Eq.\ \eqref{parameter-regime}, we find
\begin{align}
\label{pi-phase}
\mathcal R_{H/V}|_{\Delta_s=0}
\approx \mp 1
,\end{align}
which expresses the conditional $\pi$ phase shift of cavity Rydberg EIT.

The parameter
\begin{align}
\Lambda_V
= \sqrt{1+\frac{\Omega_c^2}{2\gamma\gamma_{rg}}}
\end{align}
is useful for describing the system at resonance $\Delta_s= 0$ because here $\nu_V = 1/\Lambda_V^2$. Typically, $\Lambda_V\gg 1$, which means that $\Lambda_V \approx \Omega_c/\sqrt{2\gamma\gamma_{rg}}$. It is a central result of Ref.\ \cite{Hegels:26:theoS} that the visibility $V$ of the HES is expected to be maximized for
\begin{align}
\label{Lambda-dn-C}
\Lambda_V
= C
\end{align}
because this minimizes the decoherence.

Most parameters of the experiment are chosen as in Ref.\ \cite{Stolz:22S}, e.g., $\ket g= \ket*{5S_{1/2},F{=}2,m_F{=}{-}2}$, $\ket e= \ket*{5P_{3/2}, \lb F{=}3, \lb m_F{=}{-}3}$, $\ket{r'}= \ket*{48S_{1/2},F{=}2,m_F{=}{-}2}$, $\ket r= \ket*{50S_{1/2},\lb F{=}2,\lb m_F{=}{-}2}$, $\omega_\text{cav}= \omega_{eg}$, $\Delta_c= 0$, $1-\eta_\text{esc}= 1.75\%$, $(g,\kappa,\gamma)/2\pi= \unit{(1.0,2.3,3.0)}{MHz}$, and $C=21$. As in Ref.\ \cite{Stolz:22S}, the incoming control light pulse has a Poissonian photon number distribution. Here, its mean photon number is $\bar n_c =0.13$. Postselection upon detecting exactly one control photon mostly removes cases in which zero control photons enter the setup or in which one control photon enters and is lost. Hence, the postselected subensemble exhibits essentially the same properties that one would obtain if the incoming control light pulse was a single-photon Fock state.

There is a small difference because there are occasional multi-photon events, in which the incoming control pulse contains more than one photon. As a simple example, imagine two photons taking different paths through the setup, with one in the bypass and the other one successfully stored but not successfully retrieved. In this situation one detects only one photon and it is in state $\ket V$, despite the presence of Rydberg blockade. Hence the subensemble obtained by postselection upon finding $\ket V$ contains some contamination of events that should have been attributed to Rydberg blockade, i.e., to $\ket H$. As $\bar n_c \ll 1$, such events are rare.

In contrast to Ref.\ \cite{Stolz:22S}, the present experiment uses a variable mean number of input target photons $|\alpha_\text{in}|^2$ and it uses $\Omega_c/2\pi= \unit{44}{MHz}$. Another difference to Ref.\ \cite{Stolz:22S} is that we installed an adjustable optical attenuator in front of the fiber in the DV bypass. This attenuator is not shown in Fig.\ \ref{fig-scheme}. The attenuation is adjusted such that the DV output state has equal population in states $\ket H$ and $\ket V$ for the DV input state $\ket{\psi_\theta}$ of Eq.\ \eqref{psi-theta}. As the retrieval efficiency shows some dependence on $\alpha_\text{in}$, see Sec.\ \ref{sec-larger-cats}, the attenuation must be readjusted when changing $\alpha_\text{in}$. Alternatively, one could obtain the same DV output state without this attenuator by replacing the DV input state of Eq.\ \eqref{psi-theta} by $\cos(\varphi_0) \ket H + e^{i\theta} \sin(\varphi_0) \ket V$ with $\varphi_0$ chosen, depending on $\alpha_\text{in}$, such that the DV output state has equal population in states $\ket H$ and $\ket V$.

\subsection{Measurement of the coherence time}

\label{sec-coherence-time}

In this section, we describe a novel method for measuring the coherence time $\gamma_{rg}^{-1}$. This measurement is motivated by the fact that according to Eq.\ \eqref{Lambda-dn-C}, the visibility of the HES is expected to be maximized for $\Lambda_V= C$. Hence, determining $\Lambda_V$ and $C$ in the experiment is crucial. We determine $\Omega_c$ with the same method as in Ref.\ \cite{Stolz:22S}. To determine $C$ and $\gamma_{rg}$, however, we use a novel method. The key idea behind this method is to measure $|\mathcal R_{V}|^2$ as a function of $\Omega_c$. For very large (very small) $\Omega_c$ the resonator is strongly overcoupled (strongly undercoupled), resulting in $|\mathcal R_{V}|^2\approx 1$. Between these regimes, there is a value of $\Omega_c$ at which the resonator is critically coupled, defined by the condition $|\mathcal R_{V}|^2=0$, assuming monochromatic resonant light $\Delta_s=0$. Critical coupling is found at $\Lambda_V^2= C/(2\eta_\text{esc}-1) \approx C$, i.e, $\Omega_c \approx \sqrt{2\gamma \gamma_{rg} C}$. Hence, if $C$ and $\gamma$ are known, one can infer $\gamma_{rg}$ from the value of $\Omega_c$ at critical coupling. If only $\gamma$ is known, then $C$ and $\gamma_{rg}$ can be extracted simultaneously from a fit because both affect the functional form of $|\mathcal R_{V}(\Omega_c)|^2$. If the light is not exactly resonant $\Delta_s \neq0$, one still expects a pronounced minimum of $|\mathcal R_{V}|^2$ as a function of $\Omega_c$.

To measure the coherence time $\gamma_{rg}^{-1}$ for the time scale relevant for our experiment, we use a duration of $t_t = \unit{1}{\micro s}$ for the target signal pulse, as in the rest of this paper and in Ref.\ \cite{Stolz:22S}. This causes noticeable interaction-time broadening in the measurement.

To model this, we consider the incoming oscillating electric field $\mathcal E_\text{light}(t)= \frac12 \mathcal E_0 \lb u_t(t) e^{-i\omega_0 t}+\cc$ with carrier frequency $\omega_0>0$, complex amplitude $\mathcal E_0$, and slowly varying envelope \cite{Stolz:22S}
\begin{align}
\label{ut-t}
u_t(t)
= 
\begin{cases}
\sqrt{2/t_t} \cos(\pi t/t_t), & |t|\leq t_t/2 \\
0, & \text{otherwise},
\end{cases}
\end{align}
which is normalized to $\int_{-\infty}^\infty dt |u_t(t)|^2 =1$. A Fourier transform $\widetilde u_t(\Delta_\text{in}) = (2\pi)^{-1/2} \int_{-\infty}^\infty \lb u_t(t) \lb e^{i\Delta_\text{in} t} \lb dt$ gives
\begin{align}
\label{ut-tilde-omega}
\widetilde u_t(\Delta_\text{in})
= 2\sqrt{\pi t_t} \frac{\cos(\frac12 \Delta_\text{in}t_t)}{\pi^2-\Delta_\text{in}^2t_t^2}
.\end{align}
Now $\omega_\text{in}= \omega_0+\Delta_\text{in}$ has some distribution. We always choose the carrier frequency $\omega_0$ to match $\omega_{eg}$. Hence
\begin{align}
\label{Delta-s-Delta-in}
\Delta_s
= \Delta_\text{in}
.\end{align}
Hence, the total reflected power is proportional to $\overline{\mathcal R}\null_{V}^2$, where
\begin{align}
\label{R-HV-bar}
\overline{\mathcal R}_{H/V}
= \mp \left( \int_{-\infty}^\infty d\Delta_\text{in} |\mathcal R_{H/V} \widetilde u_t|^2 \right)^{1/2}
.\end{align}
The overall factor $\mp 1$ in this definition is merely a conventions. With this convention, the conditional $\pi$ phase shift of Eq.\ \eqref{pi-phase} is carried by the $\overline{\mathcal R}_{H/V}$, not by some overlap integrals defined below. Hence
\begin{align}
\label{R-HV-bar-limit}
\overline{\mathcal R}_{H/V} \to \mathcal R_{H/V}|_{\Delta_\text{in}{=}0}
\qfor
t_t\to\infty
.\end{align}
We solve the integral for $\overline{\mathcal R}_V$ numerically and fit this model to the experimental data with $C$ and $\gamma_{rg}$ as the only free parameters. The frequency components of $\widetilde u_t(\Delta_\text{in})$ with nonzero detuning cause a nonzero value of $\overline{\mathcal R}\null_{V}^2$ at the minimum.

\begin{figure}[tb!]
\centering
\includegraphics[width=0.95\columnwidth]{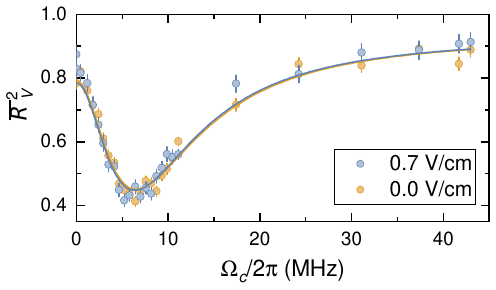}
\caption{Measurement of the coherence time. The cavity reflectivity $\overline{\mathcal R}\null_V^2$ as a function of the coupling Rabi frequency $\Omega_c$ displays a pronounced minimum at critical coupling. Fitting $\overline{\mathcal R}\null_V^2$ with $\overline{\mathcal R}_V$ of Eq.\ \eqref{R-HV-bar} to the experimental data, we obtain best-fit values for $C$ and $\gamma_{rg}$. Blue (orange) dots show experimental data with (without) applying an external electric field of \unit{0.7}{V \per cm}. The data and the best-fit curves show no discernible dependence on the electric field. The best-fit value is $\gamma_{rg}^{-1}= \unit{0.68(2)}{\micro s}$ both with and without the electric field.}
\label{fig-coherence-time}
\end{figure}

Blue dots in Fig.\ \ref{fig-coherence-time} show the experimental data in the presence of a static electric field with a strength of $\mathcal E =\unit{0.7}{V \per cm}$, as in the rest of this work and in Ref.\ \cite{Stolz:22S}. The best-fit value is $\gamma_{rg}^{-1}= \unit{0.68(2)}{\micro s}$. This result is quite different from the best-fit value \unit{0.22}{\micro s} of Appendix H of Ref.\ \cite{Stolz:22S}. This is because that reference used a method for determining $\gamma_{rg}^{-1}$, which showed only a very weak dependence on the value of $\gamma_{rg}^{-1}$. Hence, the signal-to-noise ratio in the measurement in that reference did not suffice to extract a reliable best-fit value for $\gamma_{rg}^{-1}$. In other words, the systematic uncertainty in $\gamma_{rg}^{-1}= \unit{0.22}{\micro s}$ is drastically larger than the statistical uncertainty quoted in Ref.\ \cite{Stolz:22S}.

The result of the measurement in Fig.\ \ref{fig-coherence-time} is not to be confused with the result $\gamma_{rg}^{-1}= \unit{7}{\micro s}$ of Appendix I of Ref.\ \cite{Stolz:22S} obtained from the decay of the retrieval efficiency as a function of the dark time between storage and retrieval. These numbers are expected to differ because a measurement of the decay of the retrieval \emph{efficiency} is insensitive to some sources of decoherence, in particular to electric-field noise and laser phase noise. Instead, these noise sources cause shot-to-shot fluctuations in the \emph{phase} of the retrieved light \cite{Schmidt-Eberle:20S}. The discrepancy between the values obtained from the decay of the retrieval efficiency and from Fig.\ \ref{fig-coherence-time} suggests that the value $\gamma_{rg}^{-1}= \unit{0.68(2)}{\micro s}$ is dominated by electric-field noise or laser phase noise.

\subsection{Electric-field noise}

\label{sec-electric-field}

To obtain an upper bound for the electric-field noise $\Delta \mathcal E$ from the best-fit value of $\gamma_{rg}$, we use the static electric polarizability $\alpha_r= 2\pi\hbar \times \unit{50.6}{MHz(cm\per V)^2}$ \cite{Stolz:phdS} of the Rydberg state $\ket r =\ket*{^{50}S_{1/2}}$ used here. An applied electric field $\mathcal E$ causes an energy shift of $-\frac12\alpha_r \mathcal E^2$. As the full width at half maximum (FWHM) $\Delta \mathcal E$ of the noise has the property $\Delta \mathcal E\ll |\mathcal E|$, this noise has the same effect as frequency noise of the EIT coupling laser with a FWHM of $\Delta\omega_c = |\alpha_r \mathcal E| \Delta \mathcal E/\hbar$. Assuming that the electric-field noise has a Lorentzian probability distribution, we can apply the methods of Ref.\ \cite{Gea-Banacloche:95S} to find $\Delta\omega_c \leq \gamma_{rg}$. From this, we obtain the upper bound
\begin{align}
\Delta \mathcal E
\leq \frac{\hbar\gamma_{rg}}{|\alpha_r \mathcal E|}
= \unit{7}{mV\per cm}
.\end{align}
We note that on the time scale of one hour, which is well beyond the time needed to record the data in Fig.\ \ref{fig-coherence-time}, we typically observe some drift of $\Delta_c$, which we compensate with a servo. It is unclear whether this slow drift is caused by the electric field or by the laser frequency locking setup.

To explore whether electric-field noise or laser phase noise dominates in $\gamma_{rg}$, we repeat the measurement at zero applied electric field, see the orange dots in Fig.\ \ref{fig-coherence-time}. We obtain the same best-fit value for $\gamma_{rg}^{-1}$. As the dependence of $\gamma_{rg}^{-1}$ on electric-field noise at $\mathcal E=0$ is quadratic and, hence, much weaker than at $\mathcal E =\unit{0.7}{V \per cm}$, the fact that the best-fit value for $\gamma_{rg}^{-1}$ is unchanged shows that $\gamma_{rg}^{-1}$ cannot be dominated by electric-field noise at $\mathcal E=0$.

In addition, the noise source present at $\mathcal E=0$ is most likely also present at $\mathcal E=\unit{0.7}{V/cm}$. If there was an equally large amount of noise resulting from electric-field noise at $\mathcal E=\unit{0.7}{V/cm}$, then one would expect $\gamma_{rg}$ to increase by a factor of $\approx\sqrt2$ compared to its value at $\mathcal E=0$. But such an increase is not observed. This suggests that laser phase noise is the dominant contribution to $\gamma_{rg}^{-1}$ at $\mathcal E =0$ and at $\mathcal E =\unit{0.7}{V \per cm}$.

\subsection{Incomplete Rydberg Blockade}

\label{sec-incomplete-blockade}

According to the simple model of Ref.\ \cite{Hegels:26:theoS}, the effective loss coefficient $L_g$ is expected to become as low as $L_g= 1-\eta_\text{esc} = 1.75\%$ for $\Lambda_V= C$. The best-fit value of $L_g=20.2(11)\%$ from Fig.\ \ref{fig-entangled}(b) is one order of magnitude larger, showing that the system has a large potential for future improvements. A complete analysis of the physical effects causing this discrepancy is beyond the present scope. However, we model two effects, namely incomplete Rydberg blockade and interaction-time broadening, and we study how they affect $L_g$.

First, we consider incomplete Rydberg blockade, which arises from the fact that the atomic ensemble has a nonzero radius and that the Rydberg blockade has a finite range. A model for incomplete blockade is given in Appendix C of Ref.\ \cite{Hegels:26:theoS}. Lacking a reliable value for $\gamma_{rg}^{-1}$ in the presence of EIT coupling light, Ref.\ \cite{Hegels:26:theoS} used $\gamma_{rg}^{-1}= \unit{7}{\micro s}$ in the absence of EIT coupling light from Appendix I of Ref.\ \cite{Stolz:22S}. Based thereon, incomplete blockade was expected to increase $L_g$ only very moderately to 2.1\% for $\Lambda_V = C= 21$, as stated in Ref.\ \cite{Hegels:26:theoS} in the form of $(1-L_g)/2L_g=23.4$. Applying the same methods with the value of $\gamma_{rg}^{-1}= \unit{0.68}{\micro s}$ measured here, we find a much larger value $L_g= 5.9\%$ for $\Lambda_V = C= 21$.

Obviously, increasing the coherence time $\gamma_{rg}^{-1}$ in the experiment would help. As Sec.\ \ref{sec-electric-field} suggests that $\gamma_{rg}$ is dominated by laser phase noise, one should work on that. 

Another strategy that might reduce the effect of incomplete blockade, would consist in varying $C$ and $\Omega_c$. We study this in Sec.\ \ref{sec-vary-C-Lambda}.

\subsection{Interaction-Time Broadening}

\label{sec-interaction-time}

To study interaction-time broadening, we generalize the model of Ref.\ \cite{Hegels:26:theoS}. In doing so, we assume that the \emph{spatial} mode of the spontaneously emitted light is identical for $H$ and $V$, i.e., we ignore incomplete Rydberg blockade and the finite atom number effect of Appendix B of Ref.\ \cite{Hegels:26:theoS}. First, we consider the spontaneously emitted light. Let $u_{a,H/V}(t)$ denote the normalized temporal mode function of the spontaneously emitted light, if the control polarization is $H/V$. Its Fourier transform reads
\begin{align}
\label{u-a-HV}
\widetilde u_{a,H/V}(\Delta_\text{in})
= \frac1{\overline{\mathcal A}_{H/V}} \mathcal A_{H/V}(\Delta_\text{in}) \widetilde u_t(\Delta_\text{in})
\end{align}
with a normalization constant
\begin{align}
\overline{\mathcal A}_{H/V}
= \left( \int_{-\infty}^\infty d\Delta_\text{in} |\mathcal A_{H/V} \widetilde u_t |^2 \right)^{1/2}
\end{align}
in analogy to Eq.\ \eqref{R-HV-bar}. Including interaction-time broadening, the loss coefficient $L_a$ of Eq.\ (B6) of Ref.\ \cite{Hegels:26:theoS} becomes
\begin{align}
\label{L-a}
L_a
= \frac{(\overline{\mathcal A}_H - \overline{\mathcal A}_V)^2 + 2 \overline{\mathcal A}_H \overline{\mathcal A}_V (1-\Re c_{a,H,V})}{4}
,\end{align}
where
\begin{align}
\label{c-a-HV}
c_{a,H,V}
= \ip{u_{a,H}}{u_{a,V}}
\end{align}
describes the mode overlap. Here, we use the notation $\ket{u_{\dots}}$ for normalized temporal modes. The overlap integral can be calculated in the frequency domain or in the time domain $c_{a,H,V} = \int_{-\infty}^\infty d\Delta_\text{in} \widetilde u_{a,H}^* \widetilde u_{a,V} = \int_{-\infty}^\infty dt \lb u_{a,H}^* u_{a,V}$ according to Parseval's theorem. For the light lost from the highly reflective mirrors, there is a completely analogous effect with corresponding quantities $\overline{\mathcal M}_{H/V}$, $L_m$, $c_{m,H,V}$, etc. 

For the reflected light, interaction-time broadening has a different effect. This is because the reflected temporal modes must eventually be multiplied by some normalized temporal mode function $u_h(t)$ in processing the homodyne data, as described in Sec.\ \ref{sec-homodyne}. Its Fourier transform is $\widetilde u_h(\Delta_\text{in})$. For the reflected light, we obtain normalized modes $u_{r,H/V}(t)$. Their Fourier transforms
\begin{align}
\widetilde u_{r,H/V}(\Delta_\text{in})
= \frac1{\overline{\mathcal R}_{H/V}} \mathcal R_{H/V}(\Delta_\text{in}) \widetilde u_t(\Delta_\text{in})
\end{align}
are calculated in analogy to Eq.\ \eqref{u-a-HV} using $\mathcal R_{H/V}$ and $\overline{\mathcal R}_{H/V}$ of Eqs.\ \eqref{R-HV} and \eqref{R-HV-bar}. Typically, these modes will have incomplete overlap with $\ket{u_h}$. We obtain a unique expansion in terms of the orthogonal projection onto the homodyne mode $\ket{u_h}$ and onto the subspace orthogonal thereto
\begin{align}
\label{u-r-HV-ket}
\ket*{u_{r,H/V}}
= c_{h,H/V} \ket{u_h}+ c_{\perp,H/V} \ket*{u_{\perp,H/V}}
.\end{align}
The normalized modes $\ket*{u_{\perp,H/V}}$ are found using the Gram-Schmidt process, i.e., we define expansion coefficients
\begin{align}
c_{h,H/V}
= \ip*{u_h}{u_{r,H/V}}
\end{align}
and $c_{\perp,H/V} = \sqrt{1-|c_{h,H/V}|^2}$ and the normalized modes $\ket*{u_{\perp,H/V}} = (\ket*{u_{r,H/V}} - \lb c_{h,H/V} \ket{u_h})/c_{\perp,H/V}$, where we assumed $c_{\perp,H/V}\neq 0$. The light in the modes $\ket{u_{\perp,H}}$ and $\ket{u_{\perp,V}}$ is not included in the data analysis and must be modeled as photon loss. The decomposition \eqref{u-r-HV-ket} can be modeled as a beam splitter \cite{Hegels:26:theoS}. Hence, the reflected light in a coherent state $\ket{r_V}$ with amplitude $r_V= \overline{\mathcal R}_V \alpha_\text{in}$ becomes a tensor product of coherent states $\ket{h_V \alpha_\text{in},\xi_V \alpha_\text{in}}$, where 
\begin{align}
h_V
= c_{h,V} \overline{\mathcal R}_V
,&&
\xi_V
= c_{\perp,V} \overline{\mathcal R}_V
\end{align}
are the coefficients for the modes $\ket{u_h}$ and $\ket{u_{\perp,V}}$, respectively. Analogously, $\ket{r_H}$ with amplitude $r_H= \overline{\mathcal R}_H \alpha_\text{in}$ becomes $|h_H \alpha_\text{in},\lb \xi_H \alpha_\text{in}\rangle$ with $h_H= c_{h,H} \lb \overline{\mathcal R}_H$ and $\xi_H= c_{\perp,H} \overline{\mathcal R}_H$. Note that $\sgn(\xi_{H/V}) = \sgn(\overline{\mathcal R}_{H/V})= \mp1$ because of the conditional $\pi$ phase shift and $c_{\perp,H/V} >0$.

Hence, $\ket*{r_{H/V}}$ in Eq.\ (14) of Ref.\ \cite{Hegels:26:theoS} is to be replaced by $\ket*{h_{H/V}\alpha_\text{in},\xi_{H/V}\alpha_\text{in}}$. Taking the partial trace over the lost modes $\ket*{u_{\perp,H/V}}$, we find that $L_l= L_a+L_m$ in Eq.\ (28a) of Ref.\ \cite{Hegels:26:theoS} must be replaced by $L_l = L_a+L_m+L_r$, where in analogy to Eqs.\ \eqref{L-a} and \eqref{c-a-HV}
\begin{align}
L_r
= \frac{(\xi_H - \xi_V)^2 + 2 \xi_H \xi_V (1-\Re c_{\perp})}4
\end{align}
with $c_{\perp}= \ip{u_{\perp,H}}{u_{\perp,V}}$. Using the Cauchy-Schwarz inequality, we find $|c_\perp|^2\leq 1$. Using the definition of $\ket*{u_{\perp,H/V}}$, we find
\begin{align}
c_{\perp}
= \frac{c_{r,H,V}-c_{h,H}^*c_{h,V}}{\sqrt{(1-|c_{h,H}|^2)(1-|c_{h,V}|^2)}}
\end{align}
with
\begin{align}
c_{r,H,V}
= \ip{u_{r,H}}{u_{r,V}}
\end{align}
in analogy to Eq.\ \eqref{c-a-HV}. From here, we find $L_g= \frac{L_l}{1-L_\text{cav}+L_l}$, where $L_\text{cav}= 1 - \frac14|h_V-h_H|^2$ in analog to Ref.\ \cite{Hegels:26:theoS}.

We cannot move on to calculating numbers without making a decision about the choice of the temporal homodyne mode $u_h(t)$. Obviously, it is advantageous to choose $u_h(t)$ such that $L_g$ is minimized. Aiming for an analytic solution, we prefer to minimize $L_r$ instead of $L_g$. It turns out that this makes little difference because $L_\text{cav}$ varies much more slowly than $L_r$ when varying $u_h(t)$. 

Using $2|\xi_H\xi_V|\leq |\xi_H|^2+|\xi_V|^2$ and $|c_\perp|\leq 1$ and $\xi_{H/V}\in\mathbb R$, we find the upper bound $L_r \leq \frac12(\xi_H^2+\xi_V^2)$. Hence, minimizing $\xi_H^2+\xi_V^2$ is a fairly good approximation to minimizing $L_r$. For the parameters of our experiment $\overline{\mathcal R}_V \approx -\overline{\mathcal R}_H$. Hence, we assume $\overline{\mathcal R}_V = -\overline{\mathcal R}_H$ and we find $L_r \leq \frac12(\xi_H^2+\xi_V^2) = \frac12 \overline{\mathcal R}\null_V^2 (2-F_h)$, where $F_h= |c_{h,H}|^2+|c_{h,V}|^2= \ev{\rho_1}{u_h}$ and  $\rho_1= \dyad{u_{r,H}} + \dyad{u_{r,V}}$.

Hence, we want to maximize $F_h$, which could be interpreted as the fidelity between the pure state $\ket{u_h}$ and the mixed state $\rho_1$, if the latter was normalized. Obviously, to maximize $F_h$, $\ket{u_h}$ must be an element of the 2D subspace spanned by the linearly independent sequence $(\ket{u_{r,H}},\ket{u_{r,V}})$. This subspace is obviously invariant under applying $\rho_1$. We restrict the considerations to this subspace and choose a matrix representation with respect to the basis $(\ket{u_{r,H}},\ket{u_{r,V}})$. The metric tensor reads $\smqty(1& c_{r,H,V} \\ c_{r,H,V}^* & 1)$. We find that the matrix representation of $\rho_1$ equals the metric tensor. As $\rho_1$ is Hermitian, $F_h$ is maximized if and only if $\ket{u_h}$ is an eigenvector of $\rho_1$ corresponding to its maximum eigenvalue. This eigenvector is $\propto \ket{u_{r,H}} + e^{-i\varphi_1} \ket{u_{r,V}}$ with $\varphi_1= \arg(c_{r,H,V})$.

Using Eq.\ \eqref{Delta-all-identical} and assuming complete blockade, we find $\mathcal R_{H/V}(-\Delta_s) = [\mathcal R_{H/V}(\Delta_s)]^*$. Combining this with $\widetilde u_t(-\Delta_s) = \widetilde u_t(\Delta_s) = [\widetilde u_t(\Delta_s)]^*$ from Eqs.\ \eqref{ut-tilde-omega} and \eqref{Delta-s-Delta-in}, we find $\widetilde u_{r,H/V}(-\Delta_s) \lb = [\widetilde u_{r,H/V}(\Delta_s)]^*$. Hence $c_{r,H,V}$ is always real. In the limit of infinite pulse duration $t_t\to\infty$, we find $c_{r,H,V}\to 1$ because of Eq.\ \eqref{R-HV-bar-limit}. As $c_{r,H,V}$ is a continuous function of $t_t$, we find $\arg(c_{r,H,V})= 0$ for pulses, which are not too short. For such pulse durations, the optimal homodyne mode function is
\begin{align} 
\label{u-h-superposition}
u_h(t)
= \frac{u_{r,H}(t) + u_{r,V}(t)}{\sqrt{2+2\Re\ip{u_{r,H}}{u_{r,V}}}}
.\end{align}

To simplify the result \eqref{u-h-superposition} even further, we note that for the parameters of our experiment $u_{r,V}(t)$ is approximately a delayed version of $u_t(t)$ and that $u_{r,H}(t)$ is well approximated by $u_t(t)$. To see this, we recall Eqs.\ \eqref{parameter-regime} and \eqref{Delta-all-identical}. Approximating $\mathcal R_V(\Delta_s)$ by its first-order Taylor polynomial, we find $\mathcal R_V \approx 1 + i \tau_g \Delta_s + O(\Delta_s^2)$, where we abbreviated
\begin{align}
\label{tau-g}
\tau_g 
\approx \frac2\kappa \left(1 + \frac{4\kappa\gamma C}{\Omega_c^2} \right)
.\end{align}
We rewrite this as $\mathcal R_V (\Delta_s) \approx \texp(i \tau_g\Delta_s) + O(\Delta_s^2)$, which corresponds to $u_{r,V}(t)\approx u_t(t-\tau_g)$ so that $\tau_g$ is the group delay of cavity EIT \cite{Stolz:phdS}. An analogous consideration in the presence of Rydberg blockade gives us $\mathcal R_H \approx -\texp(i \tau_{g,H} \Delta_s) + O(\Delta_s^2)$ with $\tau_{g,H} \approx \frac2C(\frac1{C\kappa}-\frac1\gamma) = \unit{-0.005}{\micro s}$ for $C=21$. This delay is negligible. We use these results to define the homodyne mode function chosen in our data analysis as
\begin{align} 
\label{u-h-final}
u_h(t)
= \frac{u_t(t) + u_t(t-\tau_g)}{\sqrt{2+2\Re\ip{u_t(t)}{u_t(t-\tau_g)}}}
\end{align}
with the incoming mode function $u_t(t)$ of Eq.\ \eqref{ut-t} and the group delay $\tau_g$. Compared to Eq.\ \eqref{u-h-superposition}, the mode function \eqref{u-h-final} has the advantage that besides the accurately known incoming mode $u_t(t)$ it contains only one free parameter $\tau_g$.

We find that the analytic approximation \eqref{tau-g} for the group delay $\tau_g$ is fairly accurate for the parameters of our experiment. For $C=21$ and $\Omega_c/2\pi=\unit{44}{MHz}$ Eq.\ \eqref{tau-g} gives $\tau_g= \unit{0.18}{\micro s}$. For comparison, a fit of $u_t(t-\tau_g)$ to numerical results for $u_{r,V}(t)$ gives the best-fit value $\tau_g=\unit{0.17}{\micro s}$ and a fidelity of 99.7\% between these two modes.

\begin{table}[!b]
\centering
\caption{Results of numerical calculations including interaction-time broadening for complete blockade, $C = \Lambda_V = 21$, $\gamma_{rg}^{-1}= \unit{0.68}{\micro s}$, and $t_t= \unit{1}{\micro s}$. Results for monochromatic light, $t_t=\infty$, are shown as a reference.}\label{tab-interaction-time-boradening}
\begin{tabular*}{\columnwidth}{c@{\extracolsep\fill}ccccc}
\hline \hline
$t_t$ & $L_a$ & $L_m$ & $L_r$ & $L_\text{cav}$ & $L_g$ \\
\hline
\unit{1}{\micro s} & 2.7\% & 1.3\% & 0.2\% & 29\% & 5.6\% \\
$\infty$ & 0 & 1.4\% & 0 & 20\% & 1.75\% \\
\hline \hline
\end{tabular*}
\end{table}

To take experimental imperfections into account, we determine the group delay from our experimental data. To this end, we first reconstruct the mode functions $u_{r,H}(t)$ and $u_{r,V}(t)$ from experimental data using the methods of Ref.\ \cite{Morin:13S} and then fit $u_t(t-\tau_g)$. In this way, we find $\tau_g= \unit{0.15(1)}{\micro s}$ for cavity Rydberg EIT and confirm the negligible group delay in the presence of Rydberg blockade. We use the latter value of $\tau_g$ for processing homodyne data throughout this work. We consider the fidelity between the function $u_t(t)$ and the mode function $u_{r,H}(t)$ reconstructed from the experiment as well as the fidelity between the fit function $u_t(t-\tau_g)$ and the mode function $u_{r,V}(t)$ reconstructed from the experiment. We find that both of these fidelities are above 99\% for all data in this work.

As an aside, if $|t - \frac12 \tau_g|\leq \frac12 (t_t -\tau_g)$, then $u_h(t)$ of Eq.\ \eqref{u-h-final} with $u_t(t)$ of Eq.\ \eqref{ut-t} is a sum of two sinusoids, which by virtue of a trigonometric identity add up to a single sinusoid $u_h(t) \propto \cos[\frac\pi{t_t} (t-\frac12 \tau_g)]$. This means that for these times $u_h(t) \propto u_t(t-\frac12 \tau_g)$. If $\tau_g\ll t_t$, the last expression holds for almost all times, because it also holds where $u_h(t)$ and $u_t(t-\tau_g)$ both vanish. Hence, we can approximate
\begin{align} 
\label{u-h-approx}
u_h(t)
\approx u_t \left(t-\frac12\tau_g \right)
\qif
\tau_g \ll t_t
.\end{align}
Obviously, this is properly normalized. For $\tau_g/t_t = 0.15$, the fidelity between Eqs.\ \eqref{u-h-final} and \eqref{u-h-approx} is 99.7\%. Hence, we might as well use Eq.\ \eqref{u-h-approx} instead of Eq.\ \eqref{u-h-final} for our experimental data taken at these parameters, including all data in Fig.\ \ref{fig-entangled}. However, for the below discussion of varying $C$ and $\Omega_c$, we find that $\tau_g/t_t$ sometimes becomes much larger and the fidelity between Eqs.\ \eqref{u-h-final} and \eqref{u-h-approx} decreases noticeably.

Now that we decided which homodyne mode function $u_h(t)$ to use, we can move on to calculating numbers. Results of numerical calculations using Eq.\ \eqref{u-h-final} are shown in Table \ref{tab-interaction-time-boradening}. In retrospect, the fact that $L_r\ll L_a+L_m$ in this table justifies all approximations used to arrive at the choice Eq.\ \eqref{u-h-final} when aiming at small $L_r$. Interaction-time broadening is found to have little effect on $L_m$. However, it increases $L_a$ and $L_\text{cav}$, both of which contribute to increasing $L_g$.

An obvious way to reduce the effect of interaction-time broadening is to increase the interaction time. This would require changing the length of the optical fiber in the DV bypass.

\subsection{Varying $C$ and $\Omega_c$}

\label{sec-vary-C-Lambda}

\begin{table}[!b]
\centering
\caption{Results $L_g$ of numerical calculations including incomplete blockade for $\gamma_{rg}^{-1}= \unit{0.68}{\micro s}$ and monochromatic light.}\label{tab-incomplete-blockade}
\begin{tabular*}{\columnwidth}{c@{\extracolsep\fill}ccc}
\hline \hline
$L_g$ & $\Lambda_V=10$ & $\Lambda_V=21$ & $\Lambda_V=37$ \\
\hline
$C=10$ & 3.8\% & 9.6\% & 21\% \\
$C=21$ & 3.0\% & 5.9\% & 12\% \\
$C=37$ & 2.7\% & 4.4\% & 8.1\% \\
\hline \hline
\end{tabular*}
\end{table}

\begin{table}[!b]
\centering
\caption{Results $L_g$ of numerical calculations including interaction-time broadening for complete blockade, $\gamma_{rg}^{-1}= \unit{0.68}{\micro s}$, and $t_t= \unit{1}{\micro s}$.}\label{tab-interaction-time-boradening-scan}
\begin{tabular*}{\columnwidth}{c@{\extracolsep\fill}ccc}
\hline \hline
$L_g$ & $\Lambda_V=10$ & $\Lambda_V=21$ & $\Lambda_V=37$ \\
\hline
$C=10$ & 10\% & 8.6\% & 9.2\% \\
$C=21$ & 12\% & 5.6\% & 5.1\% \\
$C=37$ & 20\% & 6.8\% & 4.1\% \\
\hline \hline
\end{tabular*}
\end{table}

As already alluded to, one can vary the effect of incomplete blockade by varying $C$ and $\Omega_c$. This is because the Rydberg blockade radius \cite{Stolz:22S} is an increasing function of $C/\Omega_c^2$ and larger blockade radius reduces the effect of incomplete blockade. Hence, we naively expect $L_g$ to be a decreasing (increasing) function of $C$ ($\Lambda_V$). These trends are clearly seen in the results of numerical calculations in Table \ref{tab-incomplete-blockade}. 
On a more quantitative level, $C/\Lambda_V^2$ is approximately identical for $(C,\Lambda_V)=(10,10)$ and $(C,\Lambda_V)=(37,21)$ and the corresponding values $L_g=3.8\%$ and $L_g=4.4\%$ agree within a factor of $\approx 0.86$. In contrast to the calculation of incomplete blockade in Ref.\ \cite{Hegels:26:theoS}, we find that $L_g$ is typically increased considerably by incomplete blockade. Hence, the simple idea that the minimum of $L_g$ as a function of $\Lambda_V$ might still be roughly at $\Lambda_V=C$ (because it was there for complete blockade) no longer holds. For reference, combining the values $\Omega_c/2\pi= \unit{44}{MHz}$ and $\gamma_{rg}^{-1}= \unit{0.68}{\micro s}$ gives $\Lambda_V= 37$.

Interaction-time broadening also depends on the choice of $C$ and $\Omega_c$. Results of a corresponding numerical calculation are shown in Table \ref{tab-interaction-time-boradening-scan}. When varying $C$ at fixed $\Lambda_V$, we find the lowest values of $L_g$ for $C=\Lambda_V$. On the other hand, when varying $\Lambda_V$ at fixed $C$ we find the lowest value of $L_g$ for maximum $\Lambda_V$, unless $C=10$. Hence, $L_g$ is a monotonic function of neither $C$ nor $\Lambda_V$. A naive argument aiming only at minimizing $\tau_g$ of Eq.\ \eqref{tau-g} would suggest that $L_g$ should be an increasing (decreasing) function of $C$ ($\Lambda_V$). This agrees with the numerical results as a function of $\Lambda_V$, unless $C=10$. However, the numerical results do not support this expectation for the dependence on $C$. An explanation of the physical origin of this behavior is beyond the present scope. Overall, the lowest value of $L_g$ in Table \ref{tab-interaction-time-boradening-scan} is found for $C=\Lambda_V= 37$. The numerical results of Tables \ref{tab-incomplete-blockade} and \ref{tab-interaction-time-boradening-scan} together suggest that the effects of incomplete blockade and interaction-time broadening could be minimized simultaneously by operating at large $C$, while some trade-off will be needed for the optimal choice of $\Lambda_V$. 

Of course, one would like to choose $C$ and $\Lambda_V$ such that $L_g$ is minimized. In the experiment, we varied $C$ and $\Lambda_V$ in steps of factors of $\approx 2$, similar to Tables \ref{tab-incomplete-blockade} and \ref{tab-interaction-time-boradening-scan}. We found the lowest value of $L_g$ for $C=21$ and $\Lambda_V=37$. Naively adding up the entries of Tables \ref{tab-incomplete-blockade} and \ref{tab-interaction-time-boradening-scan} for $C=21$ and $\Lambda_V=37$ would give $L_g=17\%$, which is close to the measured value $L_g=20\%$. However, this naive addition would suggest that one should obtain the considerably smaller value $L_g=11.5\%$ for $C=\Lambda_V=21$ but that is not the case in the experiment. This shows that there are additional mechanisms contributing to $L_g$ which are not included in the present model. This might be, e.g., self-blockade of the target light.

Overall, the combination of incomplete blockade and interaction-time broadening can explain some fraction of the measured value of $L_g=20\%$. To mitigate incomplete blockade, one could work on laser phase noise or reduce the radius of the atomic ensemble. To reduce interaction-time broadening, one could use a longer target pulse duration $t_t$, which would also reduce self-blockade of the target pulse. After taking such measures, one should re-optimize $C$ and $\Lambda_V$.

\section{Larger Photon Numbers}

\label{sec-larger-cats}

Aiming at repeating the experiment at larger mean numbers of target photons $\alpha_g^2$, we identified several challenges.

(i) The fairly large value of $L_g$ will be problematic when aiming at much larger $\alpha_g^2$ because it will reduce the visibility of the generated HESs. As discussed above, one should be able to reduce $L_g$ considerably, possibly approaching the low value of Ref.\ \cite{Hegels:26:theoS}. Some promising approaches to do so were discussed above.

(ii) Loss of atoms is found experimentally to increase with target photon number. This is presumably caused by spontaneous emission, which gives rise to photon-recoil heating, which in turn leads to evaporation of atoms from the shallow optical dipole trap. As a result, we have to prepare a new atomic ensemble after fewer repetitions of the experiment than in our previous work \cite{Stolz:22S}. This reduces the number of quadrature values measured in a given time and might make it difficult to obtain enough data for large $\alpha_g^2$. This could be mitigated using other atom-trapping techniques. For example, one could prepare a reservoir of ultracold atoms, e.g., in an optical dipole trap, somewhat similar to many present-day Rydberg atom array experiments, see, e.g., Ref.\ \cite{Endres:16S}, but with bulk rather than single-atom control. From such a reservoir, one could reload some atoms into the cavity quickly many times without the time consuming preparation of an ultracold atomic ensemble starting from room temperature for each reloading step.

(iii) Self blockade of the target light increases with target photon number. This might be mitigated by increasing the pulse duration $t_t$.

(iv) The retrieval efficiency of the control photon is found experimentally to decrease with target photon number. This is probably caused by many-body decoherence \cite{Li:15:SpinWaveS, Murray:16S, Karanikolaou:phdS}. This also reduces the rate at which we can measure quadrature values. Presently, this is not a dominant limitation.

\end{document}